\documentclass[aps,prd,showpacs,tightenlines,nofootinbib,floatfix,preprint,superscriptaddress]{revtex4-1}
\pdfoutput=1

\usepackage{epsfig}
\usepackage{graphicx}
\usepackage{multirow}
\usepackage[inline]{enumitem}
\usepackage{amsmath,amssymb}
\usepackage[export]{adjustbox} 
\usepackage{tabularx}
\usepackage[pdfencoding=auto, psdextra]{hyperref}
\usepackage[T1]{fontenc}
\usepackage{aas_macros}
\usepackage{upgreek}
\usepackage[pscoord]{eso-pic}
\newcommand{\diff}[1]{d {#1} \, }

\newcommand{\be}{\begin{equation}}
\newcommand{\ee}{\end{equation}}
\newcommand{\bea}{\begin{eqnarray}}
\newcommand{\eea}{\end{eqnarray}}

\newcommand{\cmg}{\mathrm{cm}^2/\mathrm{g}}
\newcommand{\Msun}{M_\odot}
\newcommand{\sigvm}{\langle \sigma v\rangle /m}

\newcommand{\kms}{\mathrm{km/s}}
\newcommand{\conc}{c}

\newcommand\york{Department of Physics and Astronomy, York University,\\Toronto, Ontario, M3J 1P3, Canada}
\newcommand\uci{Department of Physics and Astronomy, University of California, Irvine, California 92697, USA}
\newcommand\durham{Institute for Computational Cosmology, Durham University, South Road, Durham, DH1 3LE, UK}
\newcommand\aachen{Institute for Theoretical Particle Physics and Cosmology~(TTK), RWTH Aachen University, D-52056 Aachen, Germany}

\begin{document}
\preprint{TTK-20-16}
\title{Velocity-dependent Self-interacting Dark Matter
from Groups and Clusters of Galaxies}

\author{Laura~Sagunski}
\email{laura.sagunski@physik.rwth-aachen.de}
\affiliation{\aachen}
\affiliation{\york}
\author{Sophia~Gad-Nasr}
\email{sophia.nasr@uci.edu}
\affiliation{\uci}
\author{Brian~Colquhoun}
\email{bcolqu@yorku.ca}
\affiliation{\york}
\author{Andrew Robertson}
\email{andrew.robertson@durham.ac.uk}
\affiliation{\durham}
\author{Sean~Tulin}
\email{stulin@yorku.ca}
\affiliation{\york}

\date{\today}

\begin{abstract}
We probe the self-interactions of dark matter using observational data of relaxed galaxy groups and clusters. 
Our analysis uses the Jeans formalism
and considers a wider range of systematic effects than in previous work, including adiabatic
contraction and stellar anisotropy, to robustly constrain the self-interaction cross section.
For both groups and clusters, our results show a mild preference for a nonzero cross section compared with cold collisionless dark matter. 
Our groups result, $\sigma/m=0.5\pm0.2\,\cmg$, 
places the first constraint on self-interacting dark matter (SIDM) at an intermediate scale between galaxies and massive clusters.
Our clusters result is $\sigma/m=0.19\pm0.09\,\cmg$, with an upper limit of $\sigma / m < 0.35\,\cmg$ ($95\%$ CL). 
Thus, our results disfavor a velocity-independent cross section of order $1\,\cmg$ or larger needed to address small scale structure problems in galaxies, but are consistent with a velocity-dependent cross section that decreases with increasing scattering velocity. 
Comparing the cross sections with and without the effect of adiabatic contraction, we find that adiabatic contraction produces slightly larger values for our data sample, but they are consistent at the $1\sigma$ level. 
Finally, to validate our approach, we apply our Jeans analysis to a sample of mock data generated from SIDM-plus-baryons simulations with $\sigma/m = 1\,\cmg$.
This is the first test of the Jeans model at the level of stellar and lensing observables directly measured from simulations.
We find our analysis gives a robust determination of the cross section, as well as consistently inferring the true baryon and dark matter density profiles.
\end{abstract}

\maketitle
\vfill
\section{Introduction \label{sec:intro}} 

Dark matter halos of galaxy groups and clusters are the most massive virialized structures in the Universe, exerting a gravitational pull on the motions of stars, gas, and light. 
Observations, in turn, have exploited these tracers to infer the existence~\cite{1933AcHPh...6..110Z,1936ApJ....83...23S}, distribution~\cite{1972ApJ...175..627R}, and microphysics of dark matter halos.
The Bullet Cluster, for example, shows that two cluster halos, measured through gravitational lensing, have merged and passed through each other, unlike the collisional gas~\cite{Clowe:2006eq}.
The interpretation is that dark matter particles do not have a large cross section for self-interactions or might even be collisionless~\cite{Randall:2007ph}.
Indeed, the paradigm of collisionless cold dark matter (CDM) has given an extraordinarily successful description of cosmic structure formation, from the recombination epoch~\cite{Ade:2015xua} to the present large-scale structure~\cite{Alam:2016hwk}.

The CDM paradigm is less successful on galactic scales.
CDM-only simulations predict cuspy halo profiles that scale approximately as $\rho_\textrm{CDM} \propto r^{-1}$ in their inner regions~\cite{Dubinski:1991bm,Navarro:1995iw,Navarro:1996gj}.
This prediction, however, has not been universally confirmed in rotation curves~\cite{Flores:1994gz,Moore:1994yx}.
Many dark matter-dominated galaxies have shallower inner profiles and lower central densities than predicted~\cite{McGaugh:1998tq,deBlok:2001hbg,Oh:2010ea}.
Baryonic feedback is one possible mechanism for reducing the central densities of halos~\cite{Navarro:1996bv,Governato:2009bg}.
However, galaxies exhibit considerable diversity in their rotation curves~\cite{deNaray:2009xj,Oman:2015xda} and no feedback model to date has explained the full scatter~\cite{Oman:2015xda,2018MNRAS.473.4392S,Kaplinghat:2019dhn}.
At the same time, there are unexpected empirical relations for rotation curves, such as a uniform halo surface density~\cite{Kormendy:2004se,Donato:2009ab} and the radial acceleration relation~\cite{Lelli:2017vgz}, which point toward a common principle underlying this diversity.
Similar tensions with CDM have arisen for the Milky Way dwarf satellites as well~\cite{Kleyna:2003zt,Walker:2011zu,BoylanKolchin:2011de}, although feedback~\cite{Wetzel:2016wro} and modeling uncertainties~\cite{Strigari:2014yea} may play a role.

Motivated by these issues, self-interacting dark matter (SIDM) has emerged as a competitor to collisionless CDM~\cite{Spergel:1999mh} (see \cite{Tulin:2017ara} for a review).
Elastic scattering between dark matter particles leads to heat transport and thermalization in the inner halo.
For low surface brightness galaxies, this has the usual effect of producing cored halo profiles with reduced central densities.
On the other hand, for high surface brightness galaxies, the inner halo may be cuspy in response to the gravitational potential from baryons~\cite{Kaplinghat:2013xca}.
This link between baryons and dark matter is a prediction of SIDM~\cite{Kaplinghat:2013xca,Vogelsberger:2014pda,Kaplinghat:2015aga,Creasey:2016jaq,Elbert:2016dbb,Robertson:2017mgj} and, remarkably, the model gives a consistent fit across the diversity of observed rotation curves for scattering cross sections per unit mass $\sigma/m \gtrsim 1\,\cmg$~\cite{Kamada:2016euw,Ren:2018jpt}.
For SIDM, this diversity simply reflects the scatter in the surface brightness of galaxies (in addition to assembly history, as for CDM)~\cite{Kamada:2016euw,Creasey:2016jaq}.
The same values of $\sigma/m$ can explain the structure and stellar kinematics of the Milky Way dwarfs as well~\cite{Zavala:2012us,Valli:2017ktb}.
SIDM also preserves the success of CDM for large scale structure since self-scattering does not affect the evolution of linear perturbations (see~\cite{Cyr-Racine:2015ihg} for a nice discussion).

In this work, we turn to galaxy groups and clusters to provide a complementary test of the SIDM model.
The Bullet Cluster, often cited as the strongest constraint on self-interactions, has a published limit of $\sigma/m < 1.25\,\cmg$ at $68\%$ confidence level (CL)  due to a null dark matter-galaxy offset~\cite{Randall:2007ph}.
However, recent simulations have shown that such offsets are smaller than previously expected~\cite{Kahlhoefer:2013dca,Robertson:2016xjh,Kim:2016ujt} and the Bullet Cluster offset does not exclude $2\,\cmg$~\cite{Robertson:2016xjh}.
In contrast, we show that much stronger limits can be obtained from the halo profiles of relaxed systems.
Similar arguments were made two decades ago, excluding cross sections above $0.1\,\cmg$ with strong lensing observations~\cite{MiraldaEscude:2000qt,Meneghetti:2000gm}.
However, these studies were based on SIDM-only simulations~\cite{Yoshida:2000uw} neglecting the gravitational effect of baryons from the brightest cluster galaxy (BCG) in the center of the halo.
More recently, Jeans modeling~\cite{Kaplinghat:2013xca,Kaplinghat:2015aga} and new SIDM simulations both with~\cite{Elbert:2016dbb,Robertson:2017mgj,Robertson:2018anx} and without baryons~\cite{Peter:2012jh,Brinckmann:2017uve} have greatly improved our understanding of self-interactions in clusters.

On the observational side, the combination of lensing and BCG stellar kinematics provides a lever for separating the stellar and dark matter mass profiles to determine the inner slope of the halo~\cite{Sand:2002cz,Sand:2003bp}.
An analysis of seven massive clusters ($M_{200} \sim 10^{15}\,\Msun$) by Newman et al.~\cite{Newman:2012nv,Newman:2012nw} found evidence for cored halos, which could be explained by SIDM with $\sigma/m \approx 0.1\,\cmg$~\cite{Kaplinghat:2015aga}. 
However, systematic uncertainties from the unknown BCG stellar mass-to-light ratio $\Upsilon_\star$ and velocity dispersion anisotropy $\beta$ may bias determinations of the inner halo profile~\cite{Schaller:2014gwa}.
Moreover, a similar study for ten galaxy groups ($M_{200} \sim 10^{14}\,\Msun$) found no evidence for cores~\cite{Newman:2015kzv}.  These systems appear to be fit well by Navarro-Frenk-White (NFW) profiles expected for CDM~\cite{Navarro:1995iw,Navarro:1996gj}.

These observations have important implications for particle physics models for SIDM~\cite{Tulin:2017ara}.
Since typical dark matter particle velocities increase with halo mass, measurements across different astrophysical scales probe $\sigma/m$ as a function of scattering velocity.
Contact interactions, with $\sigma/m$ constant in velocity, are not a viable explanation for rotation curves since the requisite cross section is too large on cluster scales~\cite{Kaplinghat:2015aga,Elbert:2016dbb}.
On the other hand, self-interactions mediated by long-range forces (compared to the de Broglie wavelength) are generally velocity-dependent and can provide a unified model for dark matter structure across all scales~\cite{Kaplinghat:2015aga}.

Our goal in the present work is to put constraints on the velocity-dependence of self-interactions on more robust footing.
We undertake a detailed study of the Newman et al.~\cite{Newman:2012nv,Newman:2012nw,Newman:2015kzv} samples of groups and clusters in the SIDM model in order to constrain $\sigma/m$ on these scales.
Our analysis is based on the spherical Jeans model for SIDM halos following Ref.~\cite{Kaplinghat:2015aga}, which we improve upon in a number of ways:
\begin{itemize}
\item The sample of groups in our analysis is a new astrophysical mass scale where SIDM has not been tested. 
In the simplest viable SIDM models, $\sigma/m$ falls with increasing velocity from dwarf to cluster scales~\cite{Kaplinghat:2015aga}.
At the scale of groups, we predict $\sigma/m$ to take on an intermediate value in the range $0.1$--$1\,\cmg$.
\item The unknown values of $\Upsilon_\star$ and $\beta$ for the BCG are nuisance parameters that may bias the inferred halo central density.
Our analysis treats both $\Upsilon_\star$ and $\beta$ as free parameters (with weak priors) to account for their uncertainty in our results for $\sigma/m$.
\item We generalize the Jeans model for SIDM halos to allow for adiabatic contraction (AC), due to the infall and cooling of baryons~\cite{Blumenthal:1985qy,Gnedin:2004cx,Gnedin:2011uj}, in the diffuse outer halo where dark matter is effectively collisionless (see also Ref.~\cite{Kaplinghat:2013xca}).
\item We construct a mock data sample from SIDM-plus-baryons simulations with $\sigma/m = 1\, \cmg$~\cite{Robertson:2017mgj} that is identical to the set of observables for the groups sample~\cite{Newman:2015kzv}.
We check that our Jeans-based analysis is able to reproduce the input value of $\sigma/m$, as well as other parameters for the dark matter and stellar profiles.
\end{itemize}
Our work is organized as follows. 
Sec.~\ref{sec:2} presents the key ingredients for our analysis: the Jeans model, with a generalization to include AC, and the observational data of galaxy groups and clusters. 
In Sec.~\ref{sec:numerics}, we give our numerical analysis. 
Our main results are a new constraint on $\sigma/m$ for groups and a reanalysis of $\sigma/m$ for clusters.
We also discuss the interplay of $\Upsilon_\star$ and $\beta$ in our results, as well as the effect of AC.
Sec.~\ref{sec:sims} concerns the simulations: constructing and fitting mock observables, following the same Jeans analysis as for the groups.
We conclude in Sec.~\ref{sec:concl}. 
Lastly, the appendices provide details for the computation of stellar line-of-sight velocity dispersions and consistency checks on our cluster profiles.

Throughout this paper, we assume a flat $\Lambda$CDM universe on large scales with cosmological parameters $\Omega_{m}=0.3,$ $\Omega_{\Lambda}=1-\Omega_{m},$ and $H_{0}=70\,\mathrm{km\, s^{-1}\, Mpc^{-1}}$. For brevity, we use $\log \equiv \log_{10}$.

\section{Observations and dark matter profiles\label{sec:2}} 

\subsection{Jeans model for SIDM halos \label{sec:Jeans}} 

The Jeans model is a semi-analytic approach for describing SIDM halo profiles in relaxed systems~\cite{Kaplinghat:2015aga,Kaplinghat:2013xca}. 
The effect of collisions is to drive the inner halo towards hydrostatic equilibrium, while the outer region remains effectively collisionless due to its lower density.
Jeans-based analyses for SIDM have been tested successfully against simulations in the optically-thin regime, where the mean free path is larger than the typical size of the collisional region (corresponding to $\sigma/m \lesssim 10\,\cmg$).
This includes simulations on dwarf and cluster scales, both with and without baryons~\cite{Rocha:2012jg, Elbert:2014bma, Vogelsberger:2014pda, Kaplinghat:2015aga, Ren:2018jpt, Robertson:2017mgj}.
Here we follow Ref.~\cite{Kaplinghat:2015aga}, neglecting proposed refinements~\cite{Sokolenko:2018noz} to the simple framework described below.

The boundary between the collisional and collisionless regions is approximated by the radius $r_1$ where dark matter has scattered on average once per particle per lifetime of the system,  according to the rate equation
\begin{eqnarray} \label{eq:rate}
\rho_{\rm SIDM}(r_1) \, \frac{\langle\sigma v\rangle}{m}\, t_{0}=1 \, .
\end{eqnarray}
Here, $\rho_{\rm SIDM}(r_1)$ is the dark matter density at $r_1$, $\langle\sigma v\rangle$ is the velocity-weighted self-interaction cross section, $m$ is the dark matter mass, and $t_{0}$ is the age of the system, which we set to $5\,\rm{Gyr}$ for groups and clusters.

The inner collisional region of the halo is modeled as a non-singular isothermal profile $\rho_{\rm iso}$, obtained by solving the time-independent Jeans equation
\begin{eqnarray} \label{eq:jeans}
\boldsymbol{\nabla} \big(\sigma_0^2 \, \rho_{\rm{iso}}(\mathbf{r}) \big) =-\rho_{\rm{iso}}(\mathbf{r}) \, \boldsymbol{\nabla}\Phi_{\rm{tot}}(\mathbf{r})\, .
\end{eqnarray}
The total gravitational potential $\Phi_{\rm{tot}}$ includes both dark matter and baryons.
For a fixed baryon density, the solution to Eq.~\eqref{eq:jeans} depends on two parameters: the central dark matter density $ \rho_0 = \rho_{\rm iso}(0)$ and the one-dimensional velocity dispersion $\sigma_0$, which is assumed to be isotropic and spatially uniform.
We also assume spherical symmetry in our analysis.

The full SIDM profile is a piecewise function delineated by $r_1$: 
\begin{eqnarray} \label{eq:matching}
\rho_{\rm{SIDM}}(r)=\left\{
        \begin{array}{ll}
            \rho_{\rm{iso}}(r) & \quad r < r_{1} \quad (\textrm{self-interacting}) \\
            \rho_{\rm{CDM}}(r) & \quad r > r_{1} \quad (\textrm{collisionless})
        \end{array}
    \right. \, .
\end{eqnarray}
While the inner halo is thermalized by self-scattering, the outer halo profile $\rho_{\rm CDM}$ is modeled as for CDM in the absence of collisions.
Finally, the two profiles are matched at $r_1$ assuming the density and enclosed mass are continuous.

For the outer CDM halo, it is well-known that CDM-only simulations yield halo profiles that are well described by the NFW profile
\begin{eqnarray}
\label{eq:NFW}
\rho_{\rm NFW}(r)=\frac{\rho_s}{(r/r_s)(1+r/r_s)^2} \, ,
\end{eqnarray}
with scale radius $r_s$ and density $\rho_s$~\cite{Navarro:1995iw,Navarro:1996gj}.
Eq.~\eqref{eq:NFW} is often equivalently parametrized in terms of the virial mass $M_{200}$, defined as the mass enclosed within the virial radius $r_{200}$ where the mean enclosed density is 200 times the critical density, and the concentration parameter $\conc = r_{200}/r_s$.
However, the infall of baryons into the halo may yield a cuspier profile for CDM through the process of AC.
The standard approach assumes an adiabatic invariant $M_{\rm tot}(r) r$, where $M_{\rm tot}(r)$ is the total enclosed mass within $r$, assuming particles are on circular orbits~\cite{Blumenthal:1985qy}.
Refs.~\cite{Gnedin:2004cx,Gnedin:2011uj} proposed a modified adiabatic invariant $M_{\rm tot}(\bar{r}) r$ to account for more realistic eccentric orbits, where $\bar{r}$ represents the orbit-averaged radius and is defined by
\begin{equation}\label{eq:rbar}
  \bar{r}/r_0 = A_0 \left( {r}/{r_{0}} \right)^w \, .
\end{equation}
In this work, we consider two possibilities for the outer halo for SIDM, either a pure NFW profile or an NFW profile modified by AC, following Eq.~\eqref{eq:rbar}.
For the latter, we adopt the AC parameters of Ref.~\cite{Gnedin:2011uj}: $r_0 = 0.03 \, r_{200}$ and $A_0 = 1.6$, while $w$ is allowed to vary in the range $0.6$--$1.3$.

For a contact interaction, the cross section is constant in velocity and we have 
\be \label{eq:constxs}
\frac{\sigma}{m} = \frac{1}{\langle v \rangle} \frac{\langle \sigma v \rangle}{m } \, .
\ee 
The mean relative velocity for scattering is $\langle v\rangle = 4\sigma_0/\sqrt{\pi}$ assuming dark matter particles have a Maxwell-Boltzmann distribution. 
Since it is useful to compare $\sigma/m$ in systems with different halo masses and typical velocities, we take Eq.~\eqref{eq:constxs} as a definition of $\sigma/m$ even though we allow for cross sections that may be velocity dependent.\footnote{For the case of anisotropic scattering, the cross section $\sigma$ should moreover be regarded as the appropriate angular moment of the differential cross section~\cite{Tulin:2013teo,Kahlhoefer:2013dca, Robertson:2016qef}. }

To summarize, the parameters of the Jeans model are as follows.
First, the baryon density must be fixed, entering via the gravitational potential in Eq.~\eqref{eq:jeans}.
With no AC, the SIDM profile is parametrized by $(\rho_0, \sigma_0)$ for the inner halo, $(M_{200}, \conc)$ for the outer NFW halo, and the matching radius $r_1$, which is related to $\sigma/m$ by Eq.~\eqref{eq:rate}.
The matching conditions at $r_1$ allow us to solve for ($M_{200}, \conc$) for a given ($\rho_0, \sigma_0$) (inside-out matching), or vice-versa (outside-in matching), leaving three independent parameters.
Allowing for AC, $w$ is a fourth parameter of the model.

Lastly, we mention there is a two-fold degeneracy manifesting in both SIDM simulations~\cite{Elbert:2014bma} and the Jeans model~\cite{Kaplinghat:2015aga} such that two values of $\sigma/m$ can yield similar spherically-averaged profiles. 
Physically, the small $\sigma/m$ solution corresponds to core growth, with thermal energy flowing inward to heat the inner halo, while the large $\sigma/m$ solution represents core collapse, with thermal energy flowing outward and the halo contracting~\cite{Colin:2002nk}. 
The core collapse regime for SIDM has been little explored by simulations~\cite{Kochanek:2000pi,Elbert:2014bma} and it is unknown whether the Jeans model remains valid in this regime due to the breakdown of the optically-thin assumption.
For relaxed clusters, there is little motivation to explore such large values $\sigma/m \gtrsim 10\,\cmg$ since they are excluded by halo shape constraints~\cite{Peter:2012jh,Robertson:2018anx}. 

To distinguish between core growth and core collapse in the Jeans model, we perform a test based on the potential energy difference $\Delta U = U_{\rm SIDM} - U_{\rm CDM}$.
$U_{\rm SIDM}$ is the total gravitational potential energy of SIDM and baryons, while $U_{\rm CDM}$ is the total potential energy for the corresponding CDM profile with the same ($M_{200}, \conc$).
Only the inner region $r < r_1$ contributes to $\Delta U$ since SIDM and CDM profiles are identical for $r > r_1$ by construction.
Since heat flows into the inner halo during core growth, we expect $\Delta U > 0$, while conversely for a halo undergoing core collapse, heat flows outward and we expect $\Delta U < 0$. 
Presently, we consider only core growth solutions to the Jeans model for which $\Delta U > 0$.

\subsection{Observational dataset \label{sec:data}} 

We confront the predictions of SIDM based on the Jeans model against a data sample of fifteen strong lensing systems.
All of these are relaxed systems dominated by a central early-type galaxy whose stellar line-of-sight velocity dispersion profiles have been measured with spatially-resolved spectroscopy.
They include:
\begin{itemize}
\item Eight galaxy groups\footnote{CSWA6, CSWA7, CSWA107, CSWA141, CSWA163, CSWA165 from the CASSOWARY survey~\cite{Stark:2013tea}; the Eight O'Clock Arc (EOCL)~\cite{Allam:2006iw}; and J09413-1100 from the SL2S survey~\cite{Limousin:2008ec}. We do not consider CSWA1 and CSWA164 from \cite{Newman:2015kzv} since they lack values of $M_{200}$ inferred by galaxy kinematics.} spanning $M_{200}\approx (0.5$--$3) \times 10^{14} \, \Msun$, selected from strong lensing surveys as representative of a scale intermediate between galaxies and massive clusters~\cite{Newman:2015kzv}.
Stellar dispersions within the central brightest group galaxy (BGG) constrain the inner mass profile, while velocity dispersions of the member galaxies are used to place a constraint on $M_{200}$, using an estimator derived from simulations~\cite{Munari:2013mh}.
\item Seven massive clusters\footnote{MS2137, A383, A611, A963, A2537, A2667, and A2390} spanning $M_{200} \approx (0.4 $--$ 2) \times 10^{15} \, \Msun$~\cite{Newman:2012nv,Newman:2012nw}.
Stellar dispersions for the central BCGs measure the inner mass profiles, while strong and weak lensing determine the outer mass profiles.
\end{itemize}
Here, we describe these observations in more detail and how they enter as priors and constraints on our modeling.

The baryon gravitational potential is a key ingredient to the Jeans model for SIDM, particularly in the central regions dominated by stars.
For the baryon densities, we take stellar luminosity profiles for the group and cluster central galaxies from Refs.~\cite{Newman:2012nv,Newman:2015kzv}.
In our fits, we allow for the stellar mass-to-light ratio $\Upsilon_\star$ to float individually for each system.
The values of $\Upsilon_\star$ may be predicted from stellar population synthesis (SPS) up to an unknown initial mass function (IMF).
Hence, it is standard practice to normalize $\Upsilon_\star$ to the SPS prediction $\Upsilon_\star^{\rm SPS}$ for a given IMF (we take a Salpeter IMF).
Similar systems are expected to share a common IMF and therefore have similar values of $\Upsilon_\star/\Upsilon_\star^{\rm SPS}$.
For the clusters, Newman et al.~\cite{Newman:2012nw} performed a joint fit assuming a common IMF for all seven BCGs, yielding $\log \Upsilon_\star/\Upsilon_\star^{\rm SPS} = 0.02 \pm 0.05^{+0.10}_{-0.16}$ for the sample.\footnote{
We quote values here normalized to a Salpeter IMF, instead of a Chabrier IMF taken in Ref.~\cite{Newman:2012nw}.
The first uncertainty is statistical, based on a fiducial analysis with isotropic stellar orbits ($\beta=0$).
The second uncertainties reflect a systematic shift induced from allowing for anisotropic orbits with $\beta = \mp 0.2$.}
Based on this, our analysis for clusters imposes a Gaussian prior on the BCG mass-to-light ratios centered at $\log \Upsilon_\star/\Upsilon_\star^{\rm SPS} = 0$ with a relatively conservative width of $0.3$.

For the groups, radial color gradients in the BGGs point toward stellar mass-to-light ratios that vary with (projected) radius, which we denote as $\Upupsilon_\star(R)$.
In Ref.~\cite{Newman:2015kzv}, these were parametrized as $\Upupsilon_\star(R) = \Upsilon_\star (R/0.3'')^{\nabla \Upsilon_\star}$, with a sample-averaged slope $\nabla \Upsilon_\star = -0.15 \pm 0.03$. 
In our fits for groups, we allow for nonzero $\nabla \Upsilon_\star$, using $-0.15 \pm 0.03$ as a Gaussian prior, and we take a weak flat uniform prior $-1 < \log \Upsilon_\star/\Upsilon_\star^{\rm SPS} < 1$, where here $\Upsilon_\star$ is the normalization, taken to be the mass-to-light ratio at $0.3''$ from the center of the BCG.
For the clusters, BCG color gradients are negligible~\cite{Newman:2012nv} and we consider only $\nabla\Upsilon_\star = 0$.

Next, we turn to the stellar kinematics for the central galaxies.
In our model, we compute line-of-sight dispersions $\sigma_{\rm LOS}$ in the standard way from the Jeans equation~\cite{1982MNRAS.200..361B}, allowing for a nonzero and constant anisotropy $\beta$ (see, e.g., \cite{Cappellari:2008kd}).
We account for atmospheric seeing and the finite spatial size of the radial bins and slit width, as discussed in~\cite{Sand:2003bp}.
Since we limit ourselves to a spherical analysis, whereas all these systems are elliptical to some degree~\cite{Newman:2012nv,Newman:2015kzv}, we also circularize the bin and slit geometry.
The details are discussed in Appendix~\ref{sec:losdisp}. 
The observed dispersions for our systems, as well as their seeing, slit geometries, and uncertainties $\delta \sigma_{\rm LOS}$, are taken from Refs.~\cite{Newman:2012nv,Newman:2015kzv}.
The agreement between the theoretical model and observations is assessed with the $\chi^2$ in the standard way,
\be
\chi^2_{\rm disp} = \sum_{\rm bins}
(\sigma_{\rm LOS}^{\rm th} - \sigma_{\rm LOS}^{\rm obs})^2/\delta \sigma_{\rm LOS}^2 \, ,
\ee
where the sum runs over the radial bins in the slit.

For the outer halos, the groups and clusters are treated slightly differently.
For the group-scale lenses, we assess our model by computing $\overline{\kappa_{\rm group}}$, which is the azimuthally-averaged mean convergence of the main central perturber of the group within the Einstein radius $R_{\rm Ein}$.
For the observed systems, $\overline{\kappa_{\rm group}}$ and $R_{\rm Ein}$ were obtained from a lensing model fit at the pixel-level and are given in Ref.~\cite{Newman:2015kzv}.
Note by construction $\overline{\kappa_{\rm tot}} = 1$; however, the lensing models for four groups have satellite perturbers included, which yields $\overline{\kappa_{\rm group}} < 1$.
The uncertainties $\delta \overline{\kappa_{\rm group}}$ are predominantly systematic and were estimated to be in the range $0.05$--$0.1$~\cite{Newman:2015kzv}.
We assess the agreement between our theoretical model and the observations with
\be \label{eq:lensing_chisq}
\chi^2_{\rm lens} = \big( \overline{\kappa_{\rm group}^{\rm th} } - \overline{\kappa_{\rm group}^{\rm obs} } \big)^2 / \delta \overline{\kappa_{\rm group}}^2 \, .
\ee
We neglect external convergence in our analysis.
In addition, we impose Gaussian priors on $\log M_{200}$ and $\log c$ for the outer halo in our model.
Ref.~\cite{Newman:2015kzv} used the velocity dispersions of the member galaxies in the groups, coupled with a scaling relation derived from simulations~\cite{Munari:2013mh}, to obtain constraints on $\log M_{200}$.
We take their quoted central values and standard deviations, $\log M_{200}^{\rm obs} \pm \delta \log M_{200}$, as inputs for our Gaussian prior on $\log M_{\rm 200}$ in our fits.
Also, we impose a Gaussian prior on $\log \conc$ based on the mass-concentration relation (MCR) that can be measured in CDM-only cosmological simulations~\cite{Ludlow:2013vxa,Dutton:2014xda,Rodriguez-Puebla:2016ofw}.
This constraint is motivated by the fact that structure formation for SIDM and CDM halos is the same on large scales and we wish to consider cosmologically realistic halos for SIDM.
For a given value of $M_{200}$, this prior is centered at the MCR-predicted value
\be
\label{eq:MCR}
\log c_{\rm MCR} = a + b \log\left( M_{\rm 200}/ \left( 10^{12}\,h^{-1}M_\odot\right) \right) \, ,
\ee
where $a,b$ are redshift-dependent quantities extracted from CDM-only simulations with a Planck cosmology~\cite{Dutton:2014xda}.
The scatter in the MCR is $\sim 0.1$\,dex~\cite{Dutton:2014xda,Ludlow:2013vxa}.
However, since strong lensing surveys are biased toward higher concentrations~\cite{Giocoli:2013tga}, we adopt a more conservative width of $0.15$\,dex in our prior for $\log \conc$.
The total $\chi^2$ for each group is
\be
\chi^2 = \chi^2_{\rm disp} + \chi^2_{\rm lens} + \chi^2_{\rm priors},
\ee
where $\chi^2_{\rm priors}$ includes our aforementioned priors on $\nabla \Upsilon_\star$, $\log M_{\rm 200}$, and $\log \conc$.

For the cluster-scale lenses~\cite{Newman:2012nv,Newman:2012nw}, each system typically has several multiply-imaged sources that, coupled with weak lensing, provide a much richer set of observables to compare to.
While a complete reanalysis of the lensing observations with SIDM profiles would be desirable, we defer this to future work.
Here we follow a simplified approach using the quoted values of $M_{200}$ and $M_{\rm tot}(100 \, {\rm kpc})$, the total enclosed mass within 100 kpc, obtained from mass modeling fits in Ref.~\cite{Newman:2012nv}.
We impose a constraint on the latter quantity as
\be
\chi^2_{\rm mass} = \big( \log M^{\rm th}_{\rm tot}(100 \, {\rm kpc})
- \log M_{\rm tot}^{\rm obs}(100 \, {\rm kpc}) \big)^2/
\delta \log M_{\rm tot}(100 \, {\rm kpc})^2 \, .
\ee
For $\log M_{\rm 200}$, we impose a Gaussian prior using the central values and widths quoted in~\cite{Newman:2012nv}.
We check the consistency of this approach {\it a posteriori} by comparing the projected mass profiles obtained in our fits to those found by full lens reconstruction~\cite{Newman:2012nv} (see Appendix~\ref{app:M2D_profile}).

Lastly, as for the groups, we impose a prior on $\log \conc$ using the MCR.
The total $\chi^2$ for the clusters is therefore
\be
\chi^2 = \chi^2_{\rm disp} + \chi^2_{\rm mass} + \chi^2_{\rm priors} \, ,
\ee
where $\chi^2_{\rm priors}$ includes our aforementioned priors for $\Upsilon_\star$, $\log M_{\rm 200}$, and $\log \conc$.
 
\section{Numerical results \label{sec:numerics}} 
\subsection{MCMCs \label{sec:MCMCs}}
We fit the observational data for the groups and clusters by performing Markov Chain Monte Carlo (MCMC) scans 
for four main setups, namely:
\begin{itemize}
    \item SIDM, either with or without AC.
    \item CDM, either with or without AC.
\end{itemize}
Depending on the setup, we have four to seven free parameters that we scan over:
\begin{equation}
\label{eq:freeparams}
M_{200},\;
\conc,\;
\sigvm, \;
w, \;
\Upsilon_{\star},\; 
\nabla \Upsilon_\star,\;
\beta \; .
\end{equation}
Following the discussion in Sec.~\ref{sec:data}, our choice of priors is listed in Table~\ref{tab:mcmcparams}. 
For SIDM, we adopt a uniform prior on $\sigvm$, as opposed to the logarithm of $\sigvm$. The latter leads to an improper posterior for systems consistent with $\sigma/m=0$.
Aside from positivity, we do not impose a direct constraint on $\sigvm$, but it is indirectly constrained by requiring $\Delta U > 0$ for core-growth solutions to the Jeans model. For CDM, we take $\sigvm = 0$.
For cases with AC, we adopt a flat uniform prior  $0.6<w<1.3$ and fix $A_0 = 1.6$ entering Eq.~\eqref{eq:rbar}, following Ref.~\cite{Gnedin:2011uj}. 
Without AC, we take the usual NFW profile, which can also be obtained as a limiting case in Eq.~\eqref{eq:rbar} by fixing $w=0$ and $A_0 = 1/0.03$.
We also consider the stellar velocity dispersion anisotropy $\beta$ as a free (constant) parameter in our fits, within the range $|\beta| < 0.3$.
For comparison, Schaller et al.~\cite{Schaller:2014gwa} have suggested that radially-biased orbits seen in CDM simulations, at the level of $\beta \sim 0.2$--$0.3$, could potentially bias determinations of the inner halo profile.

For SIDM, we perform outside-in matching where $(M_{200},c)$ are chosen for the outer halo and we then solve for ($\rho_0,\sigma_0$) of the isothermal inner halo via the matching conditions (see Sec.~\ref{sec:Jeans}). It is also possible to perform inside-out matching where ($\rho_0,\sigma_0$) are taken as free parameters and one then solves for ($M_{200},c$)~\cite{Kaplinghat:2015aga}.
We find, however, that scanning over $(\rho_0,\sigma_0)$ does not properly explore the range of small cross sections and there is a selection bias toward larger cross sections. We therefore only consider outside-in matching in this work.

\begin{table}[t]
	\centering
	\begin{tabular}{l@{\quad}|l@{\quad}|l}
		Parameter & Prior & Note\\
		\hline
        $M_{200}$ & $\log M_{200} \in N(\log M_{200}^{\rm obs}, \delta \log M_{200})$ &  
        Quoted values $\log M_{200}^{\rm obs} \pm \delta \log M_{200}$ for \\
        & & groups~\cite{Newman:2015kzv}, clusters~\cite{Newman:2012nv}, sims (Tab.~\ref{tab:stellar_fits}) \\[4pt]
        $c$ & $\log \conc \in N(\log c_{\rm MCR}, 0.15)$ & Eq.~\eqref{eq:MCR} \\[4pt]
        $\langle \sigma v \rangle/m$ & $\langle \sigma v \rangle/m \in U(0,\infty) $ & \\[4pt]
        $w$ & $w\in U(0.6,1.3)$ & AC only \\[4pt]
        $\Upsilon_\star$ & $\log \Upsilon_\star/\Upsilon_\star^{\rm SPS} \in U(-1,1) $ & Groups \& sims \\
        & $\log \Upsilon_\star/\Upsilon_\star^{\rm SPS} \in N(0,0.3) $ & Clusters  \\[4pt]
        $\nabla \Upsilon_\star$ & $\nabla \Upsilon_\star \in N(-0.15,0.03)$ & Groups  \\
         & $\nabla \Upsilon_\star =0 $ & Clusters \& sims  \\[4pt]
        $\beta$ & $\beta \in U(-0.3,0.3)$  & \\
		\hline
	\end{tabular}
	\caption{\it List of parameters and priors assumed for the samples of groups, clusters, and mock observations from simulations (Sec.~\ref{sec:sims}).
	$U(a,b)$ denotes a uniform prior distribution in the range $[a,b]$. 
	$N(\mu,\sigma)$ denotes a Gaussian prior distribution with mean $\mu$ and standard deviation~$\sigma$. $\Upsilon_\star^{\rm SPS}$~refers to the SPS reference values with a Salpeter IMF~\protect\cite{Newman:2012nv,Newman:2015kzv}. }
	\label{tab:mcmcparams}
\end{table}
%
\subsection{Cross sections \label{sec:crosssections}}

\begin{figure}[!t]
	\centering
	\includegraphics[width=0.49\textwidth,valign=t]{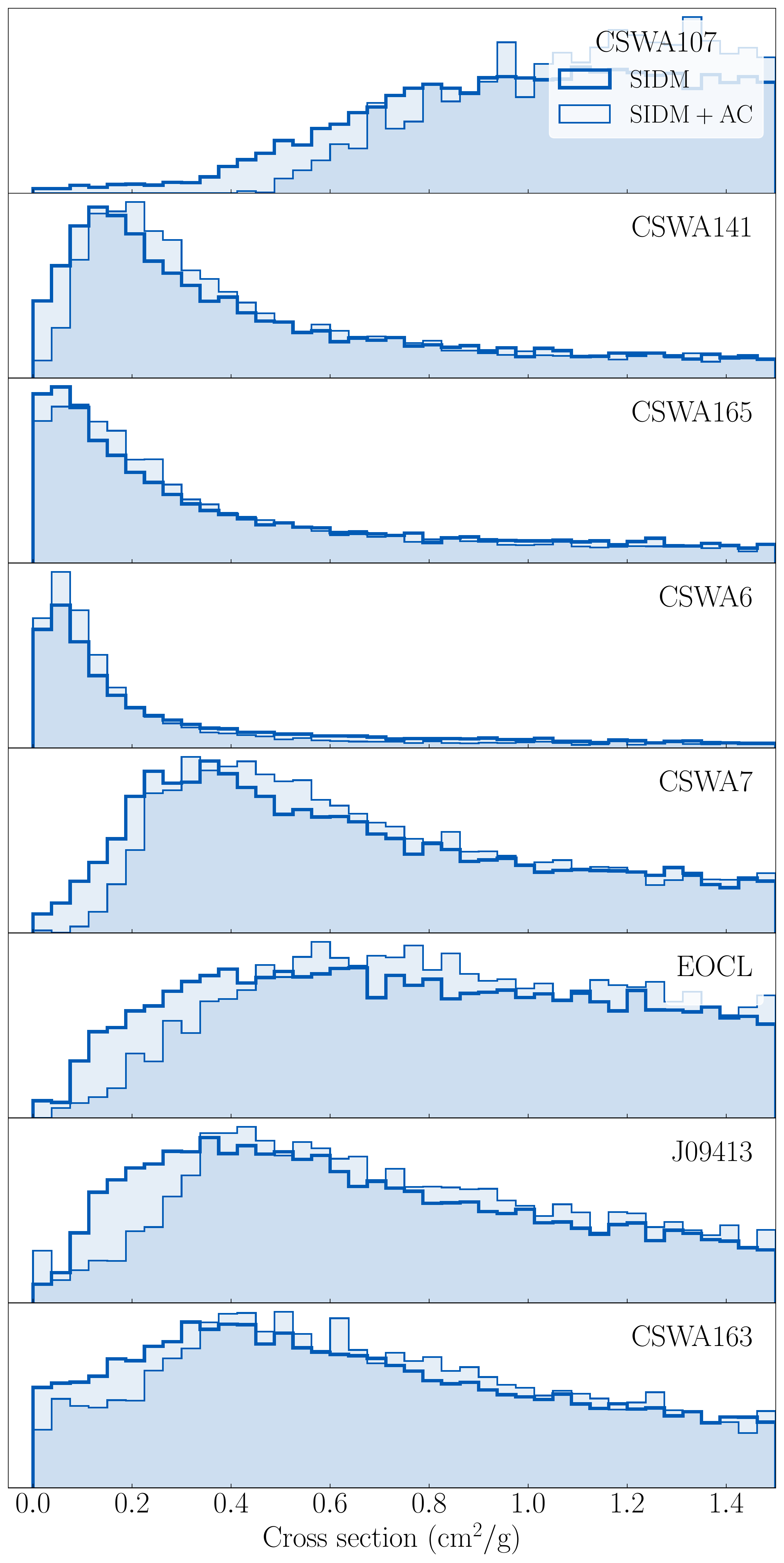}
	\includegraphics[width=0.49\textwidth,valign=t]{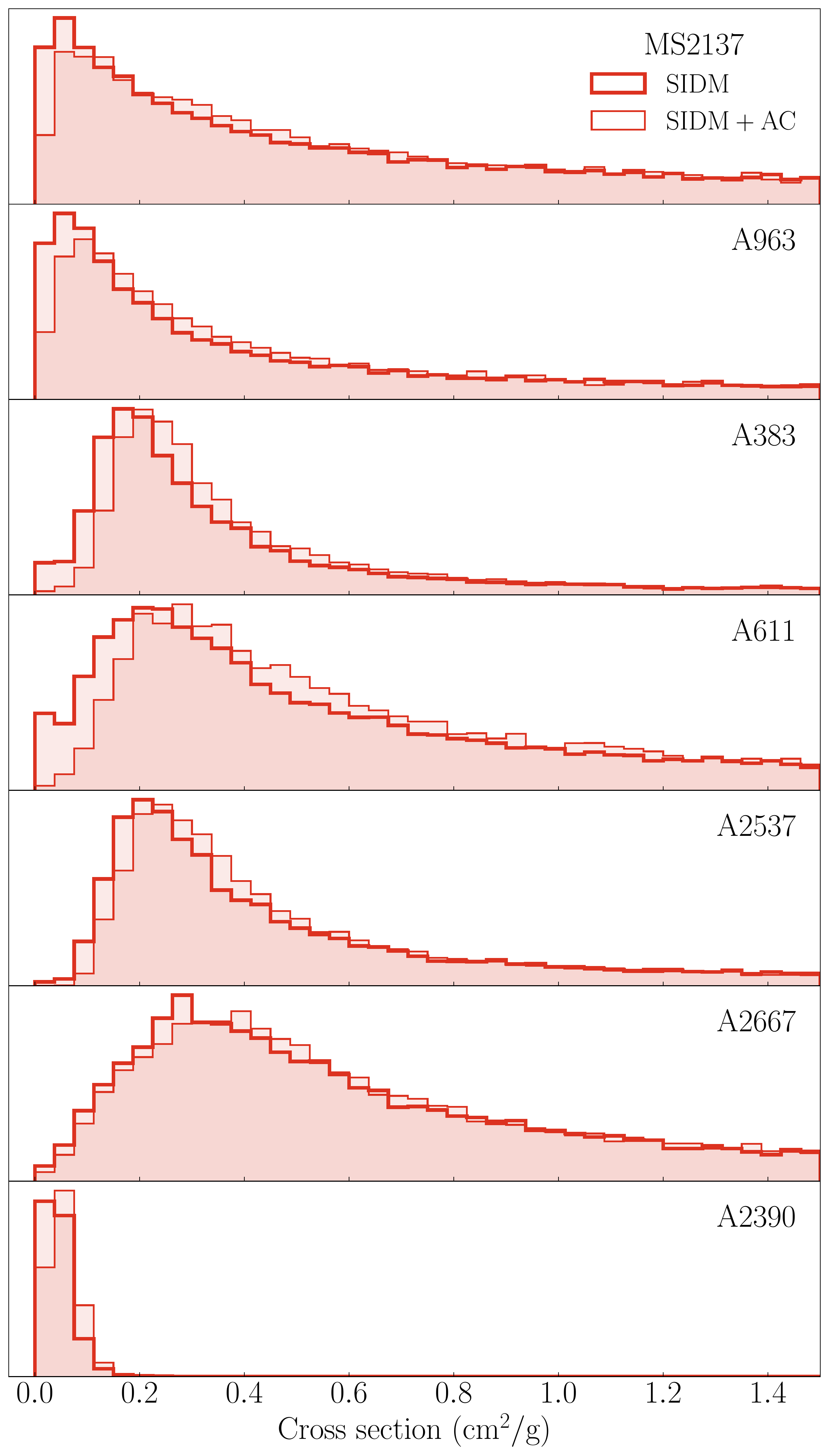}
	\caption{\it Histograms of the cross section $\sigma/m$ for the individual galaxy groups (left) and clusters~(right) in our MCMC analysis.}
	\label{fig:Histograms_groups_clusters}
\end{figure}

Here we present our main results for the self-interaction cross section.
Fig.~\ref{fig:Histograms_groups_clusters} shows the posterior distributions for $\sigma/m$ for each individual system in our analysis, both for the samples of galaxy groups (left) and clusters (right).\footnote{We calculate $\sigma/m$ from Eq.~\eqref{eq:constxs} for each point in our Markov chains.} 
Each panel compares our two prescriptions for the outer (collisionless) halo within the Jeans model.
The thick histogram is the baseline model~\cite{Kaplinghat:2015aga} where the isothermal profile is matched onto an NFW profile at $r_1$ (`SIDM'), while for the thin histogram we assume the outer NFW halo is modified by AC (`SIDM+AC').
Overall, there is a general trend that the model with AC shifts the distributions toward larger cross sections compared to the model with no AC.
Physically, this is because AC in the outer halo increases the central density in the inner halo, while larger $\sigma/m$ reduces the central density (for core-growth).
The two effects become correlated once the central density is fixed by the observations in such a way that SIDM redistributes the excess density, resulting in a larger $\sigma/m$.
We explore this effect in more detail in Sec.~\ref{sec:ACanalysis}.

In general, we see from Fig.~\ref{fig:Histograms_groups_clusters} that the preferred $\sigma/m$, indicated by the peaks of the distributions, are all below $1 \; \cmg$ (except for CSWA107).
Several groups (CSWA141, CSWA165, CSWA6) and clusters (MS2137, A963, A2390), in particular, prefer very small cross sections, $\sigma/m \lesssim 0.2\,\cmg$. In fact, the preference for a nonzero cross section compared to CDM is not very strong. Comparing the minimum $\chi^2$ values obtained by our SIDM and CDM-only fits, we find that only 3 of 8 groups and 3 of 7 clusters fulfill the criterion $\chi^2_{\mathrm{CDM}}-\chi^2_{\mathrm{SIDM}} >1$. For the remaining systems, allowing for self-interactions provides an equally good or only marginally better fit compared to CDM. Our posteriors also have large tails toward large values of $\sigma/m$, evident in Fig.~\ref{fig:Histograms_groups_clusters}, and their medians are skewed toward much larger $\sigma/m$ than their peaks. For this reason, we opt not to quote $\sigma/m$ values inferred from each system individually and simply present Fig.~\ref{fig:Histograms_groups_clusters} as is.

To quantify our results, we determine joint constraints on the cross section from our samples.
We denote $P_n(\sigma/m)$ as the probability density for each individual system $n$ shown in Fig.~\ref{fig:Histograms_groups_clusters}.
Rather than simply taking a direct product to compute a joint probability density $P(\sigma/m)$, we
recall that the rate equation~\eqref{eq:rate} depends on the product $(\sigma/m) t_0$, where the age of the system $t_0=5 \, \mathrm{Gyr}$ was fixed.
In fact, each system has its own unique age $t_n$, which yields a different inferred cross section $(\sigma/m)_n$ that is related to the true $\sigma/m$ value by a rescaling
\be
(\sigma/m)_n = (\sigma/m) t_n/t_0 \, .
\ee
With this point in mind, we construct a joint probability density for $\sigma/m$ for $N$ systems by taking a product of $P_n$ and marginalizing over each age $t_n$,
\be \label{eq:jointprob}
P(\sigma/m) \propto \prod_{n=1}^N \int_0^\infty \diff{t_n} \pi(t_n) \, P_n\big((\sigma/m)_n\big)  \, .
\ee
We adopt a Gaussian prior $\pi(t_n)$ for $t_n$ that is centered at $t_0 = 5\,\mathrm{Gyr}$, with width $\delta t = 2\,\mathrm{Gyr}$.\footnote{With a change of variables, Eq.~\eqref{eq:jointprob} becomes
\be \notag
P(\sigma/m) \propto \frac{1}{(\sigma/m)^N} \prod_{n=1}^N \int_0^\infty \diff{(\sigma/m)_n} \pi\left( \frac{(\sigma/m)_n}{\sigma/m} t_0\right) \, P_n\big((\sigma/m)_n\big)  \, .
\ee
This is straightforwardly computed from our Markov chains by
\be
\notag
P(\sigma/m) \propto \frac{1}{(\sigma/m)^N} \prod_{n=1}^N \sum_{(\sigma/m)_n \in \mathcal{C}_n} \, \pi\left( \frac{(\sigma/m)_n}{\sigma/m} t_0\right)  \, ,
\ee
where $\mathcal{C}_n$ denotes the chain of $\sigma/m$ values for each system $n$.}

\begin{figure}[!t]
	\centering
        \includegraphics[width=0.49\textwidth]{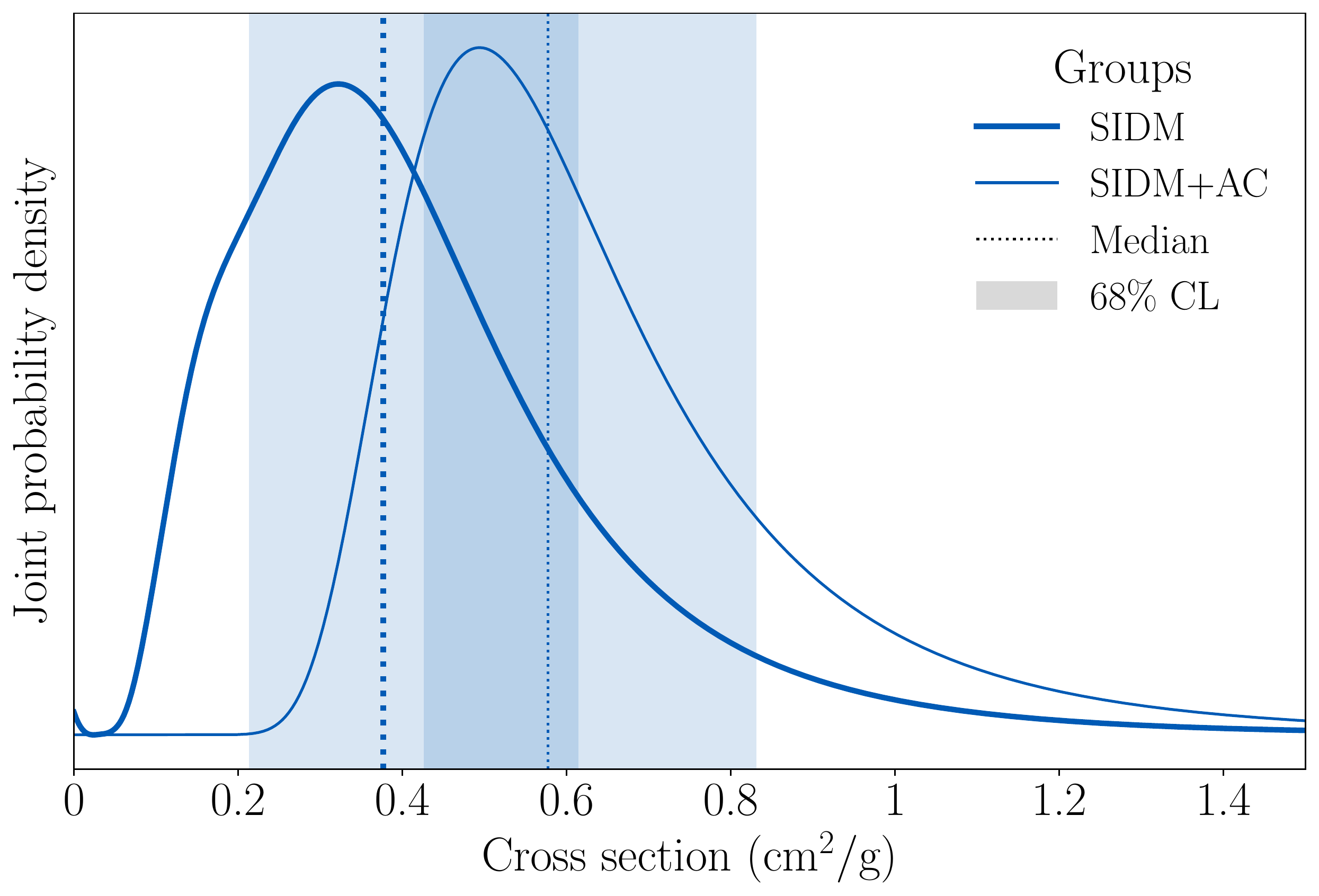}
        \includegraphics[width=0.49\textwidth]{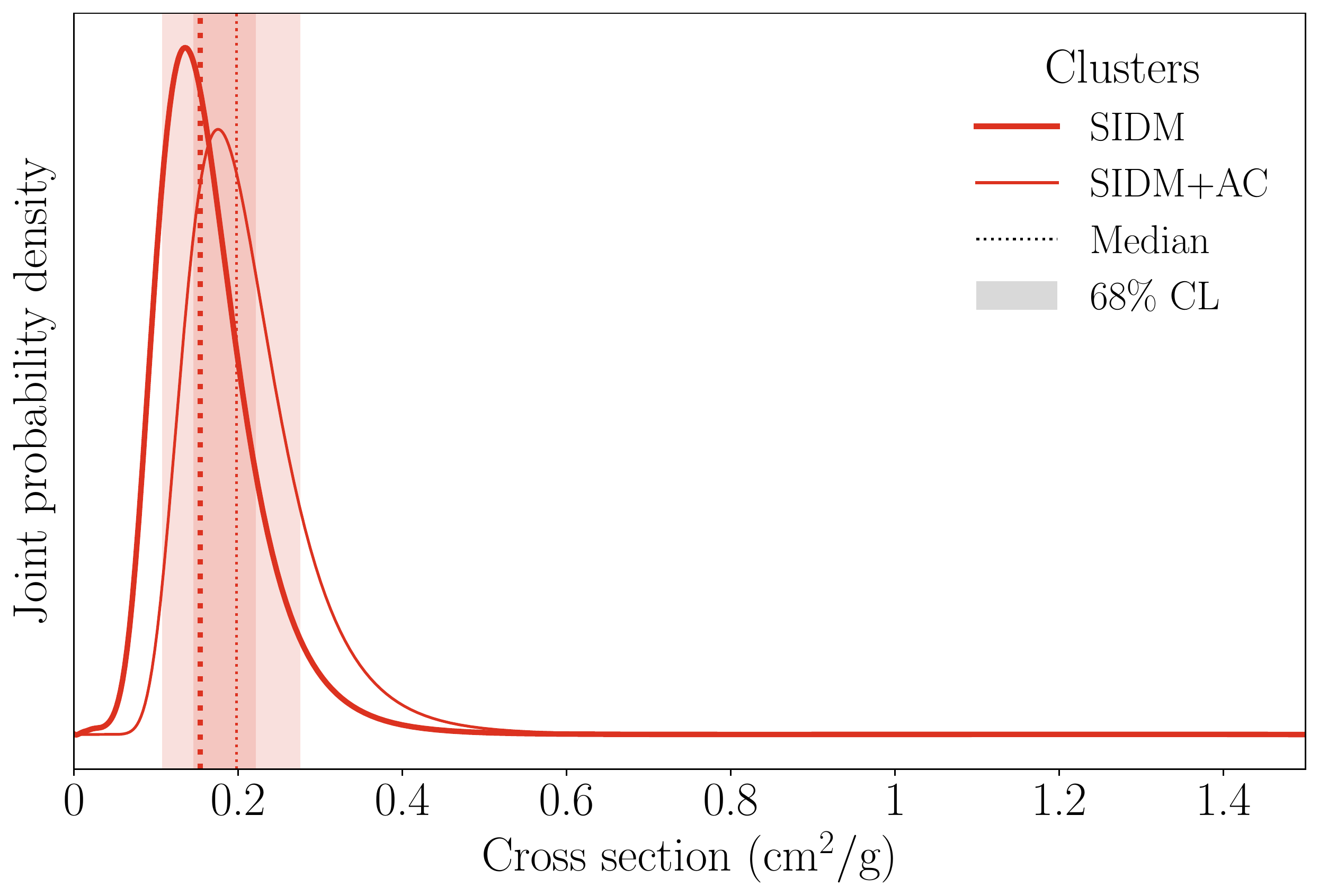}
	\caption{\it Joint probability density for $\sigma/m$ for the samples of galaxy groups~(left) and clusters (right).}
	\label{fig:PDFs_groups_clusters}
\end{figure}
Fig.~\ref{fig:PDFs_groups_clusters} presents our results for $P(\sigma/m)$, computed separately for the samples of galaxy groups (left) and clusters (right). 
The thick and thin curves represent the case without and with AC, respectively, with the latter favoring higher $\sigma/m$ values.
The dashed vertical lines indicate the median values of $\sigma/m$, while the shaded bands correspond to $68 \%$ CL intervals around the median. 
Numerically, we have for the case with NFW matching (`SIDM')
\begin{equation}
\label{eq:crosssection}
\begin{alignedat}{1}
\sigma / m &= 0.4^{\,+0.2}_{\,-0.2}\,\cmg\; \textnormal{(groups)}, \\
\sigma / m &= 0.15^{\,+0.07}_{\,-0.05}\,\cmg\; \textnormal{(clusters),}
\end{alignedat}  
\end{equation} 
and somewhat larger values for the case with AC in the outer halo (`SIDM+AC'),
\begin{equation}
\label{eq:crosssection_AC}
\begin{alignedat}{1}
\sigma / m &= 0.6^{\,+0.3}_{\,-0.2}\,\cmg\; \textnormal{(groups)}, \\
\sigma / m &= 0.20^{\,+0.08}_{\,-0.05}\,\cmg\; \textnormal{(clusters).} 
\end{alignedat}  
\end{equation} 
While allowing for AC pushes our constraints to larger values of $\sigma/m$, both cases give similar results at the $1\sigma$ level.
If we instead interpret our results as upper limits ($95\%$ CL), we have for NFW matching
\begin{equation}
\label{eq:crosssection_limit}
\begin{alignedat}{1}
\sigma / m &< 0.9\,\cmg\; \textnormal{(groups)}, \\
\sigma / m &< 0.28\,\cmg\; \textnormal{(clusters), } 
\end{alignedat}  
\end{equation} 
and slightly weaker limits for the case with AC in the outer halo,
\begin{equation}
\label{eq:crosssection_limit_AC}
\begin{alignedat}{1}
\sigma / m &< 1.1\,\cmg\; \textnormal{(groups)}, \\
\sigma / m &< 0.35\,\cmg\; \textnormal{(clusters).} 
\end{alignedat}  
\end{equation}

\begin{figure}[!t]
	\centering
	\includegraphics[width=0.95\textwidth]{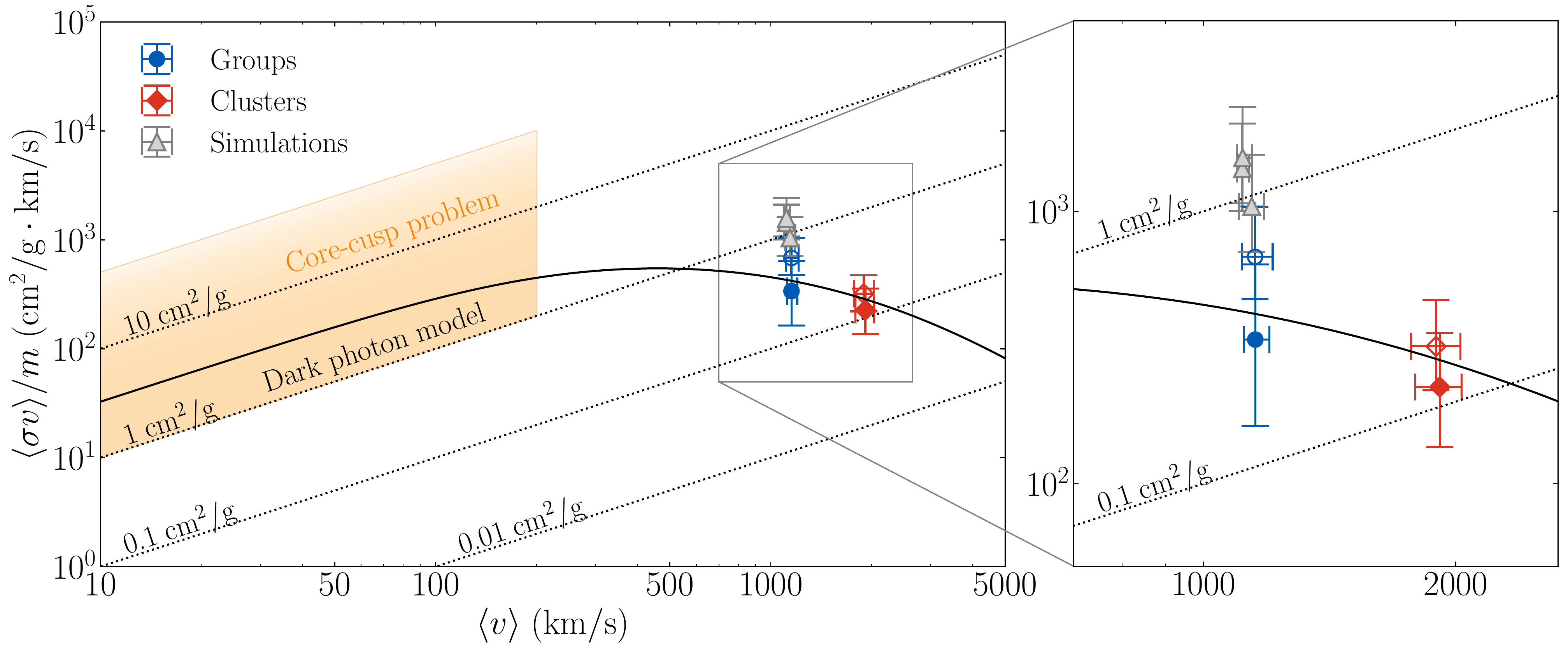} \\
	\caption{\it Velocity-dependence of self-interactions, given in terms of mean velocity-weighted cross section per unit mass $\sigvm$ versus mean scattering velocity $\langle v \rangle$. Blue (red) points are from our joint fits to  galaxy groups (clusters). Closed (open) circles correspond to SIDM fits without (with) AC. Gray points are from joint fits to mock observations from SIDM-plus-baryons simulations with $1\,\cmg$~\protect\cite{Robertson:2017mgj}, assuming different slit orientations. Shaded area is preferred SIDM range for solving core-cusp problem on dwarf scales. Solid line shows $\langle \sigma v\rangle/m$ for $15\,\mathrm{GeV}$ mass dark matter with self-interactions mediated by an $11\,\mathrm{MeV}$ dark photon, consistent with constraints across all scales.
	}
	\label{fig:sigmavm_vs_v}
\end{figure}
Our results disfavor constant cross sections at the level of $1\,\cmg$ or larger needed to address small scale structure issues in galaxies.
However, they are consistent with velocity-dependent self-interactions where $\sigma/m$ falls with increasing velocity.
Since typical velocities are larger in more massive systems, our samples constitute complementary probes for $\sigma/m$ as a function of scattering velocity.

With mean scattering velocities $\langle v \rangle \approx 1150\,\kms$ and $\approx 1900\,\kms$ for groups and clusters, respectively, our results in Eqs.~\eqref{eq:crosssection} and \eqref{eq:crosssection_AC} show a mild preference for velocity dependence.
Constructing analogous joint probability densities for $\langle \sigma v \rangle/m$, we find for the case with NFW matching
\begin{equation}
\label{eq:crosssection_velocity}
\begin{alignedat}{1}
  \langle \sigma v \rangle / m &= 340^{\,+300}_{\,-180}\,\cmg\cdot\kms\; \textnormal{(groups)}, \\
\langle \sigma v \rangle / m &=  230^{\,+130}_{\,-90}\,\cmg\cdot\kms\; \textnormal{(clusters)},
\end{alignedat}  
\end{equation} 
and for the case with AC included
\begin{equation}
\label{eq:crosssection_AC_velocity}
\begin{alignedat}{1}
\langle \sigma v \rangle / m &=  680^{\,+360}_{\,-200}\,\cmg\cdot\kms\; \textnormal{(groups)}, \\
\langle \sigma v \rangle / m &=  320^{\,+150}_{\,-100}\,\cmg\cdot\kms\; \textnormal{(clusters).}
\end{alignedat}  
\end{equation}
The $95\%$ CL upper limits we find for NFW matching are
\begin{equation}
\begin{alignedat}{1}
\langle \sigma v \rangle / m &< 950\,\cmg\cdot\kms\; \textnormal{(groups)}, \\
\langle \sigma v \rangle / m &< 480\,\cmg\cdot\kms\; \textnormal{(clusters),} 
\end{alignedat}  
\end{equation} 
whereas for the case with AC in the outer halo we obtain
\begin{equation}
\begin{alignedat}{1}
\langle \sigma v \rangle / m &< 1430\,\cmg\cdot\kms\; \textnormal{(groups)}, \\
\langle \sigma v \rangle / m &< 620\,\cmg\cdot\kms\; \textnormal{(clusters).} 
\end{alignedat}  
\end{equation}
We have rounded the values for $\langle \sigma v \rangle$ to $10\,\cmg\cdot\kms$ here.

To illustrate our results, Fig.~\ref{fig:sigmavm_vs_v} shows $\sigvm$ as a function of $\langle v\rangle$.
The blue and red circles show our joint constraints for $\sigvm$ from groups and clusters, respectively, each for cases both with (open circles) and without (closed circles) AC for the outer collisionless halo.
The dotted gray contours are lines of constant $\sigma/m$.
The shaded orange area represents the cross section range $1\,\cmg \lesssim \sigma / m \lesssim 100\,\cmg$ for SIDM to solve the core-cusp problem on galactic scales~\cite{Kaplinghat:2015aga}.\footnote{The precise upper bound is unknown. From Ref.~\cite{Elbert:2014bma}, it is likely that values beyond $\sim 100 \, \cmg$ lead to the onset of core collapse within a Hubble time.}
Lastly, the gray points show the results of our Jeans model and MCMC analysis applied to three samples of mock observations obtained from hydrodynamical SIDM simulations of clusters with $1\,\cmg$~\cite{Robertson:2017mgj}.
The fact that our analysis is able to reproduce approximately $1\,\cmg$ as an output is a reassuring test of our methods.
Sec.~\ref{sec:sims} contains discussion of these simulations and mock observations, as well their implications for our analysis. 

Due to the different velocities across astrophysical scales, the dependence of the cross section on velocity becomes apparent in Fig.~\ref{fig:sigmavm_vs_v}. 
This has important implications for particle physics models for SIDM.
Contact-type interactions are disfavored over models with light mediators that allow for velocity dependence. 
As an example of the latter, we consider dark matter self-scattering through the exchange of a dark photon, analogous to Rutherford scattering.
Self-interactions are described by a repulsive Yukawa potential $V(r) = \alpha^\prime e^{-\mu r}/r$, where $\alpha^\prime$ is the dark fine structure constant and $\mu$ is the dark photon mass~\cite{Feng:2009hw,Buckley:2009in,Loeb:2010gj,Aarssen:2012fx,Tulin:2012wi,Tulin:2013teo}. 
As an illustrative example, we take a dark matter mass $m=15\,\mathrm{GeV}$ and dark photon parameters $\alpha^\prime=1/137$ and $\mu=11\,\mathrm{MeV}$. 
The solid line in Fig.~\ref{fig:sigmavm_vs_v} shows the resulting $\sigvm$ as a function of $\langle v\rangle$, yielding a velocity dependence consistent with our new observational constraints from galaxy groups and clusters.
On dwarf galaxy scales, the typical momentum transfer $m \langle v \rangle$ is smaller than $\mu$. 
Here, self-interactions behave like a contact interaction with constant cross section $\sigma/m \approx 3 \, \cmg$, which is consistent with rotation curves~\cite{Kaplinghat:2015aga,Ren:2018jpt}.
In an analogous way, any given model of SIDM can be confronted against these observational constraints. 
This, in turn, dramatically narrows down (or excludes) the viable parameter space for all SIDM models to solve small scale structure issues.

\subsection{Adiabatic contraction \label{sec:ACanalysis}}
 
\begin{figure}[!t]
\centering
\includegraphics[width=0.6\textwidth]{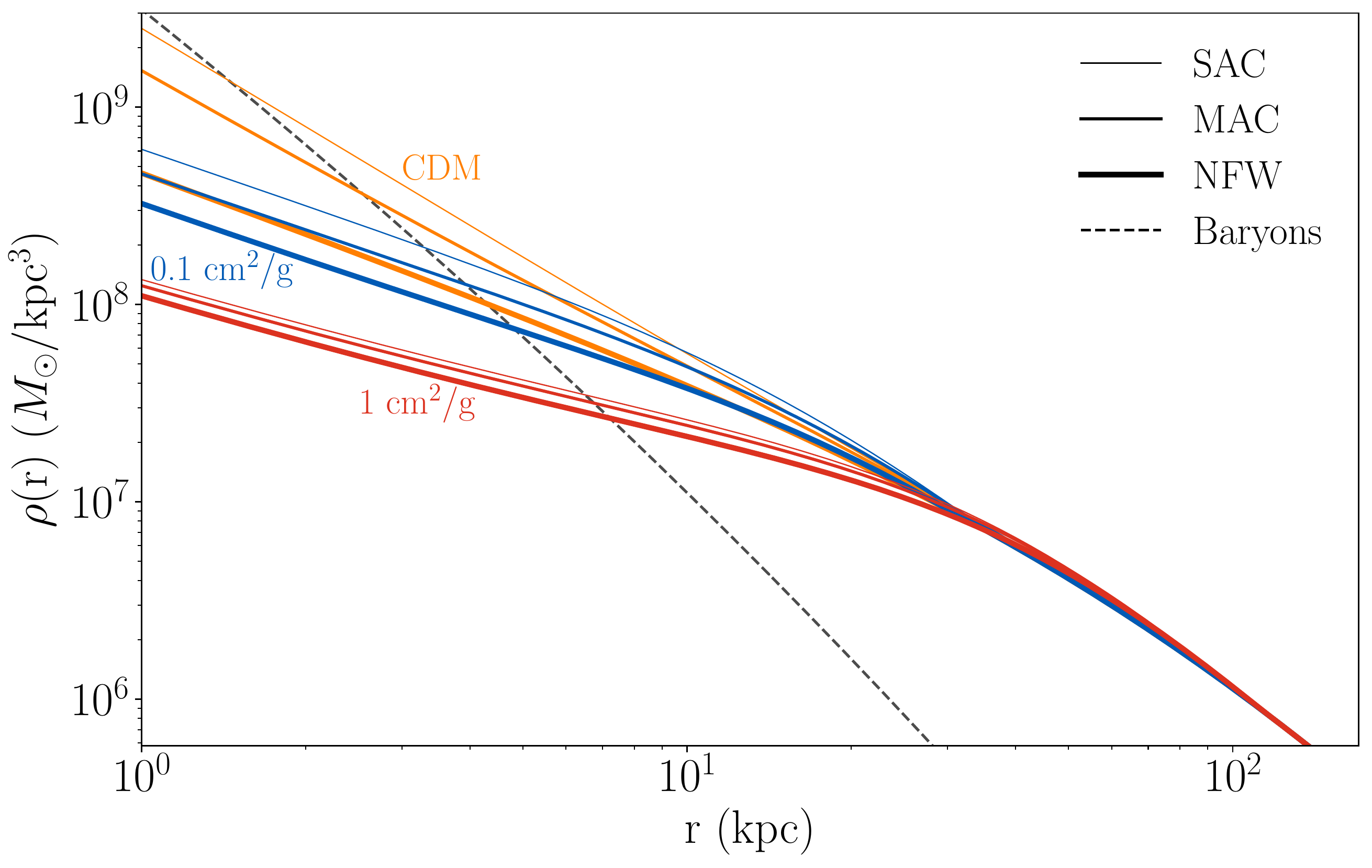}
\caption{\it Dark matter density profiles for CSWA6 for CDM and SIDM, for different AC prescriptions (see text) and cross sections, with fixed parameters $M_{200}=10^{13.84}\,M_\odot$, $\conc=10^{0.88}$, and $\Upsilon_*=10^{0.48}$, along with the baryon profile.}
\label{fig:density_plots}
\end{figure}

In our analysis of groups and clusters, we have generalized the Jeans model to allow for AC in the outer halo, beyond $r_1$ where SIDM is effectively collisionless.
Since self-interactions redistribute dark matter particles within $r_1$, it is not obvious how AC in the outer halo impacts the density profile in the inner halo for SIDM.
Here we explore the effect of AC on SIDM profiles and its impact on stellar kinematics and lensing observables.
We consider
\begin{enumerate*}[label=(\roman*),font=\itshape]
\item standard adiabatic contraction (SAC), as originally proposed in Ref.~\cite{Blumenthal:1985qy}, which assumed purely circular particle trajectories;
\item modified adiabatic contraction (MAC), proposed in Ref.~\cite{Gnedin:2004cx} to allow for elliptical trajectories; and
\item NFW profiles without AC.\footnote{In Eq.~\eqref{eq:rbar}, SAC corresponds to $A_0 = w = 1$ and MAC corresponds to $A_0 = 0.85\times0.03^{-0.2} \approx 1.7$ and $w=0.8$.
In the notation of Ref.~\cite{Gnedin:2004cx}, Eq.~\eqref{eq:rbar} was alternatively expressed as $\bar{r}/r_{200} = A \left( {r}/{r_{200}} \right)^w$ and MAC corresponds to $A=0.85$ and $w=0.8$.
The case without AC follows from setting $A=1$ and $w=0$.
}
\end{enumerate*}

For groups and clusters, the effect of AC is appreciable for SIDM halos only for small cross sections, when $r_1$ is not much larger than the radius over which baryons dominate.
We illustrate this effect in Fig.~\ref{fig:density_plots}, focusing on CSWA6, one example within the sample of groups~\cite{Newman:2015kzv}, to fix the baryon density that drives AC (dashed line).
Fixing $M_{200}$, $c$, and $\Upsilon_\star$, we show the resulting halo profiles for SIDM with cross sections of $0.1\,\cmg$ (blue) and $1\,\cmg$ (red), as well as for pure CDM (orange), each for the three AC cases.
For CDM, we have the well-known result that AC produces halos with steeper inner slopes and increased central densities compared to NFW profiles without AC~\cite{Blumenthal:1985qy,Gnedin:2004cx,Gnedin:2011uj}.
For SIDM, there remains a general trend that AC in the outer halo increases the central density in the inner halo (consistent with \cite{Kaplinghat:2013xca} for galactic scales), although the inner slopes are not steepened.
For $1\,\cmg$, the outer halo outside $r_1 \approx 115\,\mathrm{kpc}$ is far beyond where baryons are dominant so that AC has little impact, while for $0.1\,\cmg$, the effect is more appreciable since $r_1 \approx 30\,\mathrm{kpc}$.
We also note some similarity between the density profiles of CDM and SIDM with $0.1\,\cmg$ across different AC models.
This suggests some degeneracy between $\sigma/m$ and the choice of AC prescription, which explains the slightly different values of $\sigma/m$ found in Eqs.~\eqref{eq:crosssection} and \eqref{eq:crosssection_AC} for our fits without and with AC, respectively. 
This degeneracy does not persist to larger values of $\sigma/m$ as it is clear that the profiles with $1\,\cmg$ have appreciably lower central densities that are not modified significantly by AC.

\begin{figure}[!t]
\centering
\includegraphics[width=0.48\textwidth,valign=t]{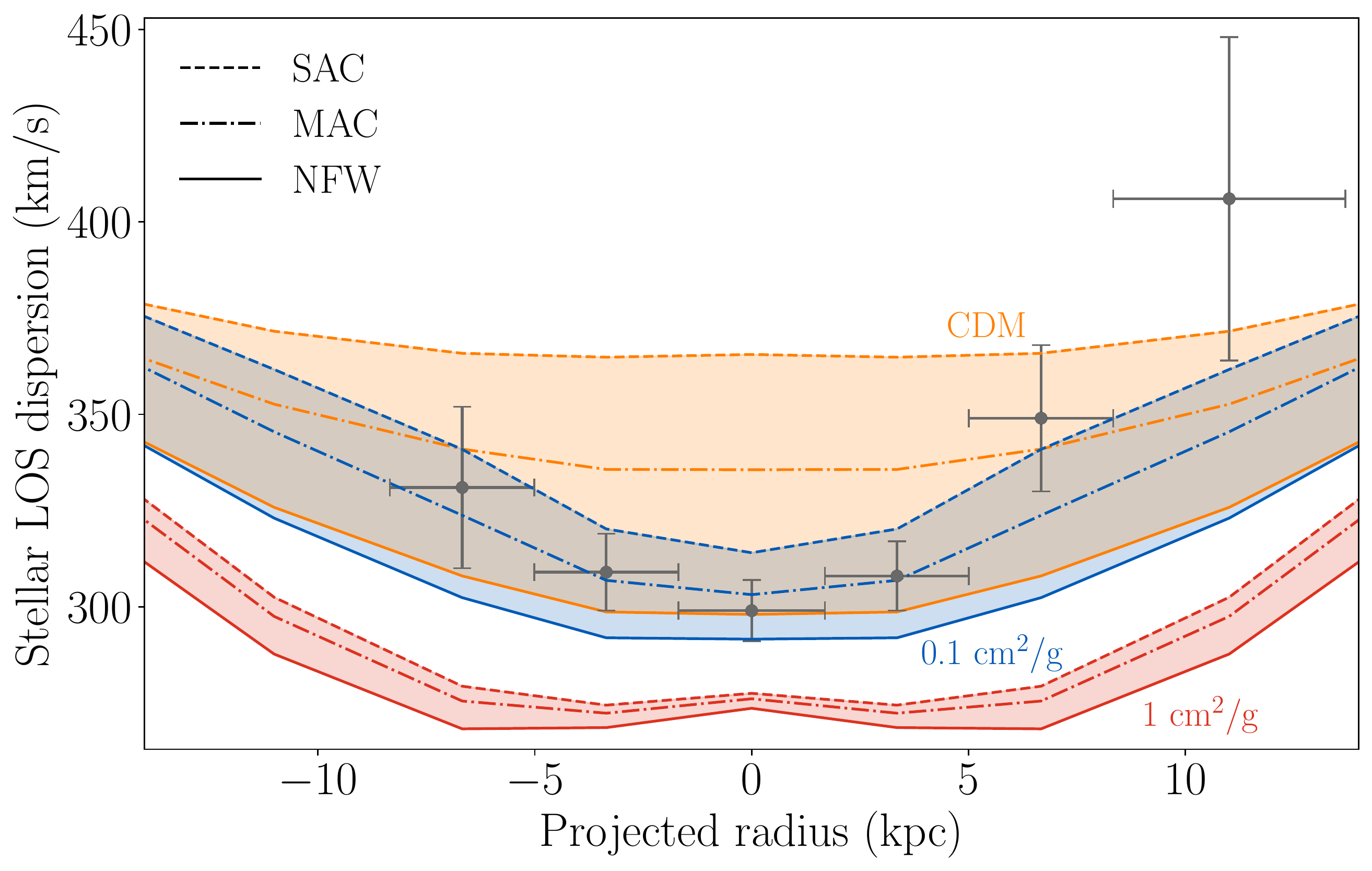}
\includegraphics[width=0.49\textwidth,valign=t]{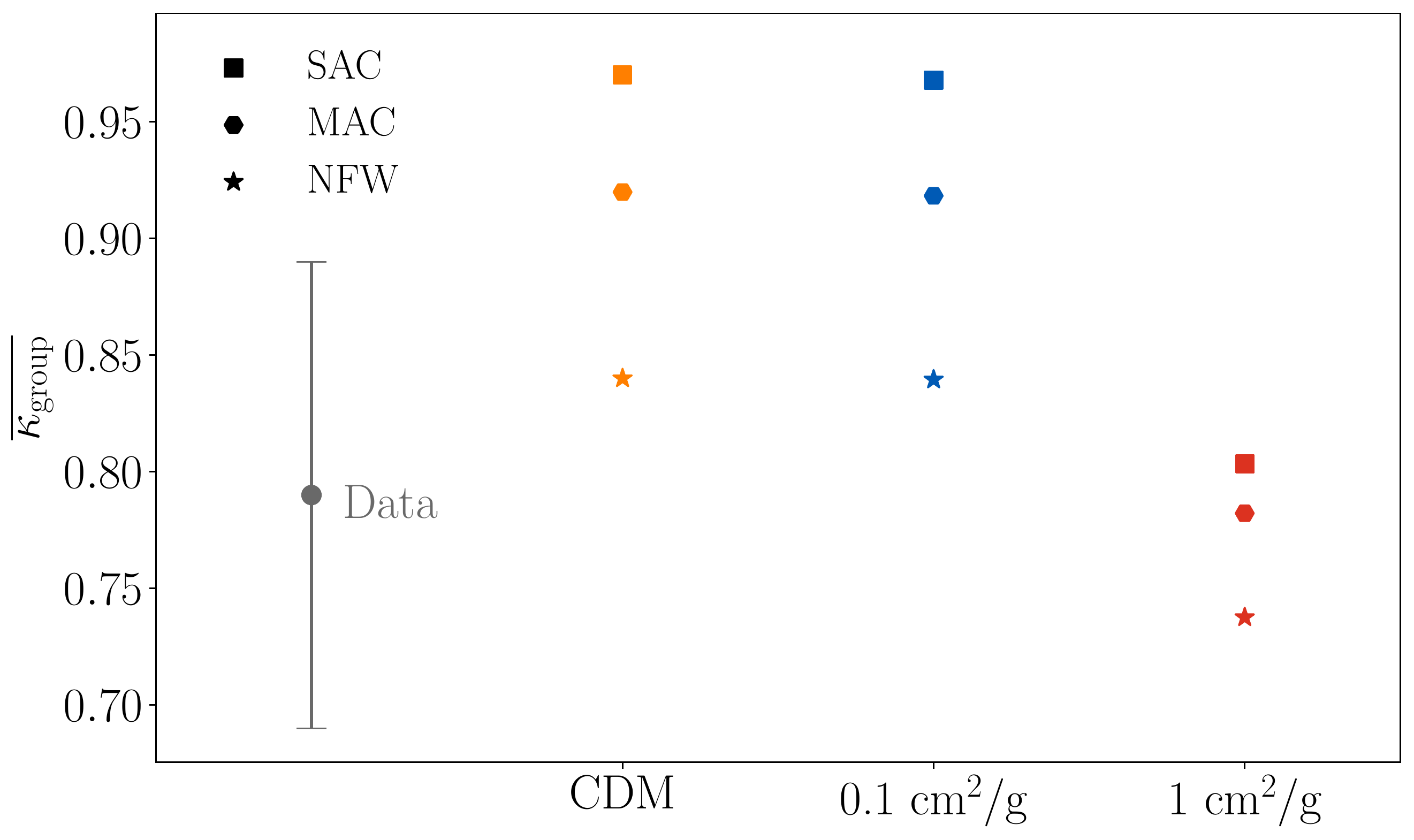}
\caption{\it Observables for CSWA6 for CDM and SIDM are shown for different AC prescriptions and cross sections, with parameters and profiles fixed as in Fig.~\ref{fig:density_plots}, compared to observations (data points).
Left panel: Stellar velocity dispersions (bands) show the spread between AC prescriptions for the cases of cross sections of $0.1\,\cmg$ and $1\,\cmg$, and pure CDM. Right panel: Mean convergence for strong lensing, $\overline{\kappa}_{\rm group}$, is plotted for CSWA6 for these AC prescriptions and cross section values.}
\label{fig:los_kappa_plots}
\end{figure}

Next, we investigate how AC directly affects the observables.
Fig.~\ref{fig:los_kappa_plots} shows the velocity dispersion profile (left) and mean group convergence $\overline{\kappa_{\mathrm{group}}}$ (right) corresponding to each of the models shown above in Fig.~\ref{fig:density_plots}.
The data points show the observed velocity dispersion and mean convergence for CSWA6.
These are intended only as a point of reference since models here have not been specifically fit to the data.
We see that the velocity dispersions show significant variation depending on the AC prescription for both CDM and (to a lesser extent) SIDM with $0.1\,\cmg$,
whereas AC has minimal impact for SIDM with $1\,\cmg$. Additionally, the overlap between the bands for CDM and SIDM with $0.1\,\cmg$ again points to a modest degeneracy between $\sigma/m$ and choice of AC for small cross sections, which does not occur for $1\,\cmg$.
The mean group convergence shows similar results. 
While AC has a minimal impact on $\overline{\kappa_{\mathrm{group}}}$ for SIDM with $1\,\cmg$, it has a larger effect for $0.1\,\cmg$ and CDM.
In fact, the latter two yield very similar values for $\overline{\kappa_{\mathrm{group}}}$ for the different AC prescriptions.
%
\subsection{Stellar mass-to-light ratio and anisotropy \label{sec:Upsilon_beta}}
There are two thorny issues for modeling the stellar component of groups and clusters.
First, the overall normalization $\Upsilon_\star$ is largely unknown due our ignorance of the IMF entering stellar population synthesis (SPS) models.
Since stars dominate the mass density in the central region, they must be modeled correctly to extract the inner halo profile.
Second, the stellar velocity dispersion anisotropy is also unknown since motions are measured only along the line of sight.
Our ignorance of anisotropy parameter $\beta$ is a systematic uncertainty in relating observed stellar kinematics to the underlying total mass density (see Appendix).
In our analysis, we have adopted relatively weak priors on both $\Upsilon_\star$ and $\beta$, allowing them to float in our fits.\footnote{We do not investigate to what extent a radial dependence of $\Upsilon_\star$ or $\beta$ (beyond including $\nabla\Upsilon_\star$ for the groups) may impact our results.} 
In this section, we investigate the preferred values of $\Upsilon_\star$ and $\beta$ that we obtain.

\begin{figure}[!t]
	\centering
	\includegraphics[width=0.49\textwidth,valign=t]{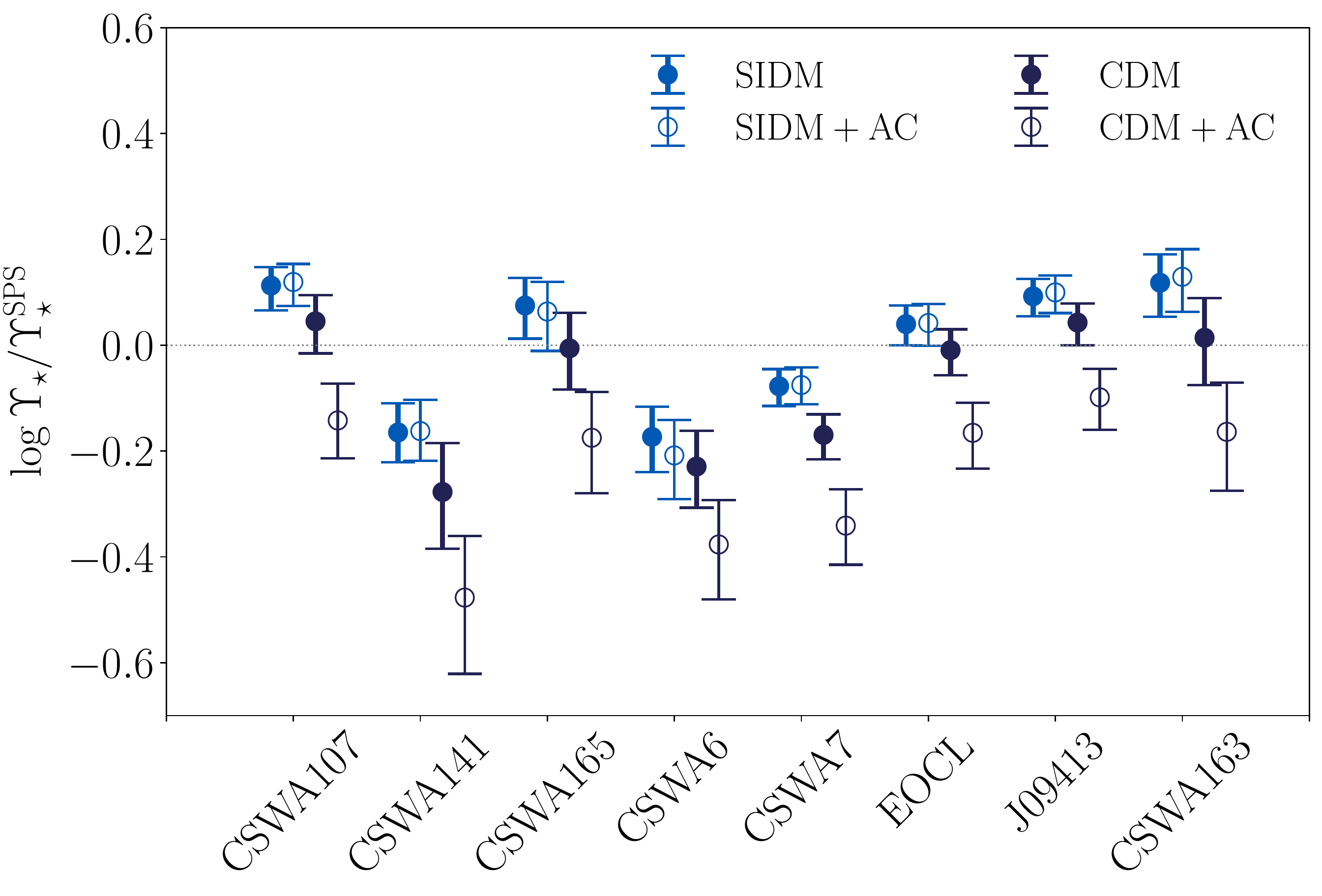}
	\includegraphics[width=0.49\textwidth,valign=t]{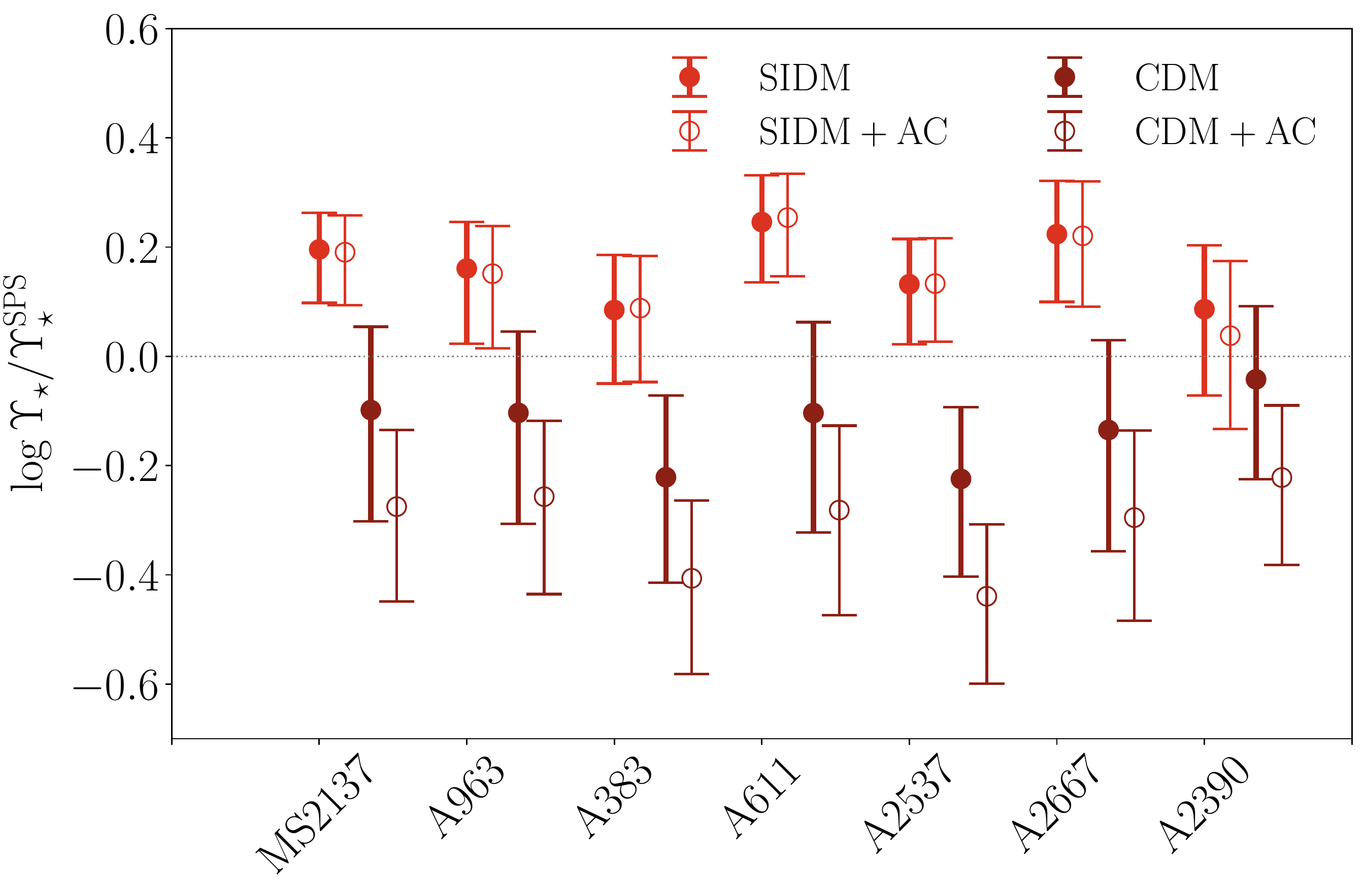}
	\caption{\it Preferred values for the mass-to-light ratio relative to the SPS values with a Salpeter IMF, $\Upsilon_\star/\Upsilon_\star^{\rm SPS}$, for the individual galaxy groups (left) and clusters (right) in our sample. Closed circles represent the preferred values for SIDM and CDM halos without AC, while open circles correspond to the cases with AC.}
	\label{fig:Bestfit_Y_Groups}
\end{figure}
Our results for $\Upsilon_\star$ are shown in Fig.~\ref{fig:Bestfit_Y_Groups} for groups (left) and clusters (right), relative to the reference values $\Upsilon_\star^{\rm SPS}$ assuming a Salpeter IMF.
For each system, we have performed four different fits: closed circles are from our CDM and SIDM fits without AC, while open circles include AC. 
For the groups as well as the clusters, we find $\Upsilon_\star/\Upsilon_\star^{\rm SPS}$ is generally larger for SIDM than for CDM. 
For CDM, our fits with AC and without AC yield preferred values for $\Upsilon_\star/\Upsilon_\star^{\rm SPS}$ that differ by $\sim 1 \sigma$.
For SIDM, interestingly, we find that $\Upsilon_\star/\Upsilon_\star^{\rm SPS}$ is basically unchanged whether AC is present or not.
However, comparing Eqs.~(\ref{eq:crosssection}) and~(\ref{eq:crosssection_AC}), it is the preferred values for $\sigma/m$ that change by $\sim 1 \sigma$ to compensate for the effect of AC. 

For groups, the combination of stellar kinematics and strong lensing in our fits is able to constrain $\Upsilon_\star$ for each system within a narrow range, despite our weak prior on $\Upsilon_\star$.
For the clusters, our results for $\Upsilon_\star$ have slightly larger error bars. In general, our SIDM fits yield values of $\Upsilon_\star$ that are larger than those in Newman et al.~\cite{Newman:2012nw,Newman:2012nv,Newman:2015kzv} by around 0.1\,dex. 
Taking a simple mean of central values in Fig.~\ref{fig:Bestfit_Y_Groups}, we obtain $\langle \log \Upsilon_\star/\Upsilon_\star^{\rm SPS} \rangle \approx 0$ for the groups and 
$\approx 0.1$
for the clusters, independent of whether or not AC is included. 
In comparison, Newman et al.\ found $\langle \log \Upsilon_\star/\Upsilon_\star^{\rm SPS} \rangle = -0.11 \pm 0.06$ for the groups~\cite{Newman:2015kzv} and 
$0.02 \pm 0.05$
for the clusters~\cite{Newman:2012nw}. 
We note, however, that our assumptions for the dark matter halo profiles are very different from Refs.~\cite{Newman:2012nw,Newman:2012nv,Newman:2015kzv}.
The latter adopted generalized NFW profiles for which the inner slope is the free parameter, whereas for SIDM, the dark matter profile is coupled to the baryon density.

Besides the cross section and the mass-to-light ratio, another important parameter in our fits is the stellar dispersion anisotropy $\beta$.
The impact of $\beta$ on inferred dark matter profiles was studied by Schaller et al.~\cite{Schaller:2014gwa}, based on a sample of clusters from CDM simulations with similar BCG surface brightness and line-of-sight velocity dispersion profiles (but typically smaller halo masses) as the Newman et al.~clusters~\cite{Newman:2012nw,Newman:2012nv}.
They argued that radially-dependent $\beta$ profiles found in their simulations, with a mean preference toward positive values $\beta \sim 0.2$--$0.3$, was inconsistent with the Newman et al.\ analysis with $\beta=0$, which in turn could bias the latter towards larger baryon densities, overestimating $\Upsilon_\star$, and inferring shallower halo profile slopes.
However, these arguments were not corroborated in more recent studies based on CDM simulations with larger halo masses~\cite{He:2019svf}.

\begin{figure}[!t]
	\centering
	\includegraphics[width=0.85\textwidth]{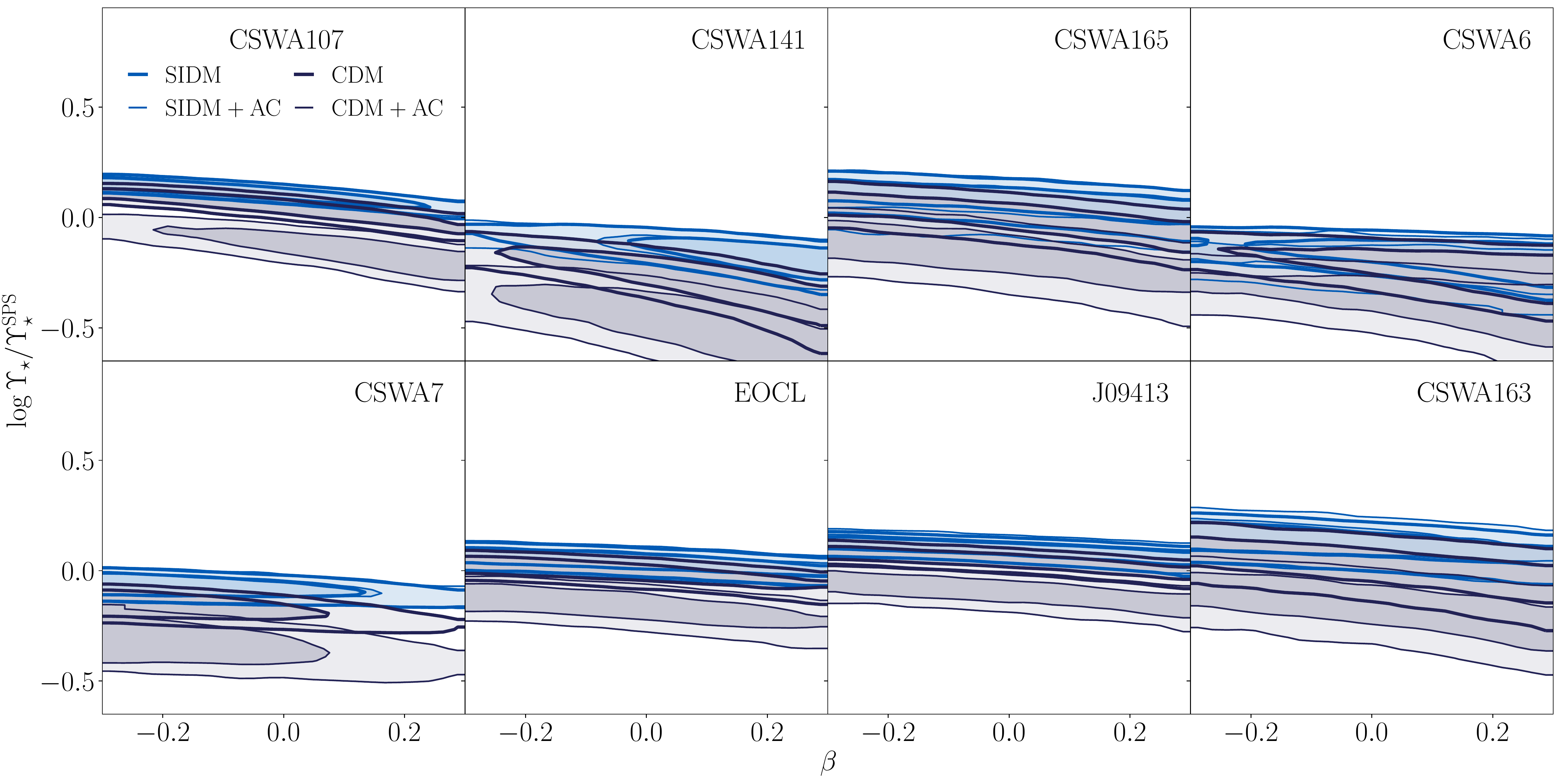}
	\includegraphics[width=0.85\textwidth]{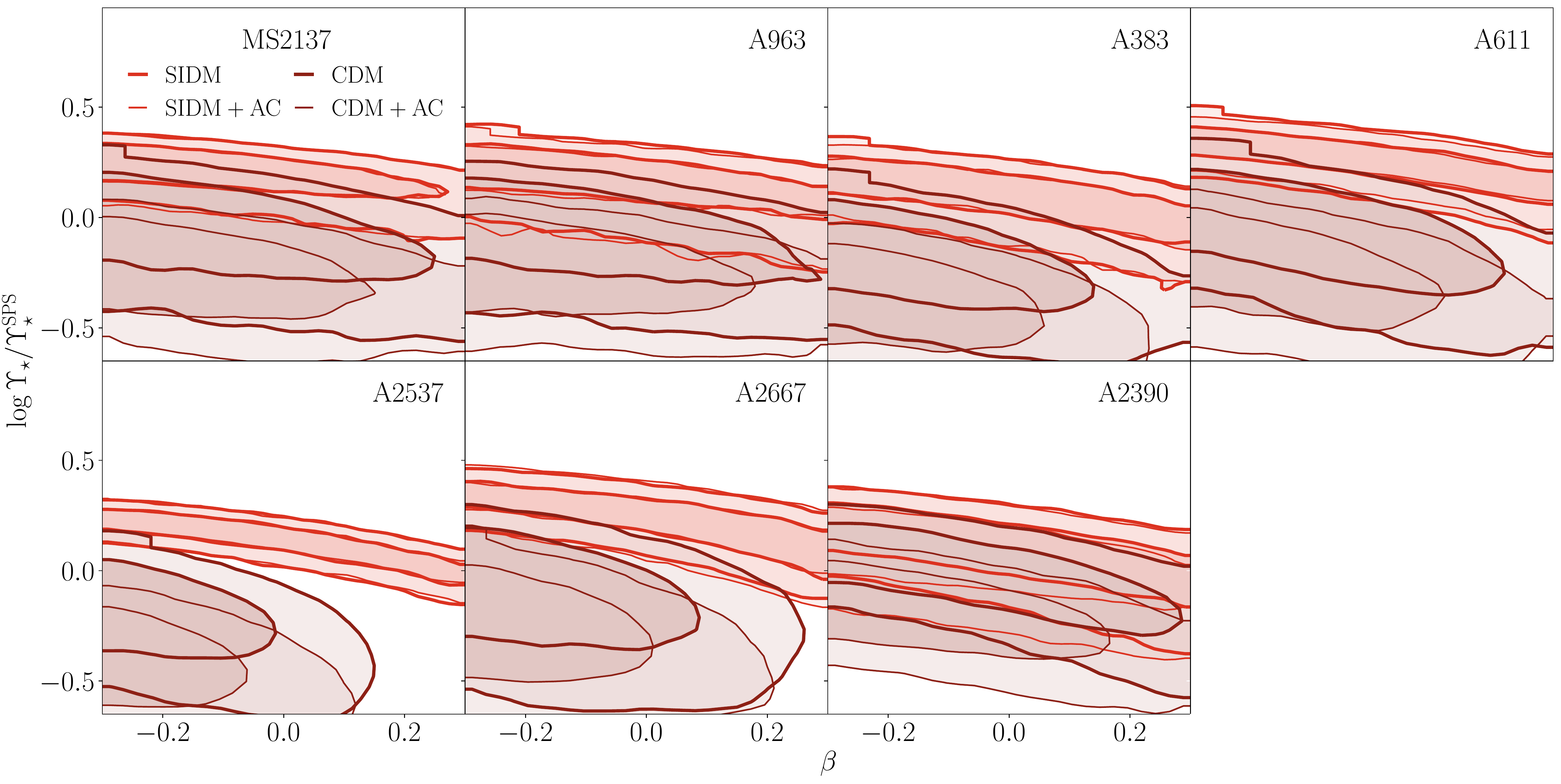}
	\caption{\it Mass-to-light ratio $\Upsilon_\star$ relative to the SPS values with a Salpeter IMF as a function of $\beta$ for the individual galaxy groups (top) and clusters (bottom). The inner and outer contours in the plots indicate $68$ and $95\%$ confidence levels of the posterior distributions for SIDM and CDM halos with and without including AC.}
	\label{fig:Y_vs_beta}
\end{figure}
Our analysis allows $\beta$ to float freely, in the range $|\beta| < 0.3$.
In Fig.~\ref{fig:Y_vs_beta}, we plot $\Upsilon_\star/\Upsilon_\star^{\rm SPS}$ as a function of $\beta$ for the individual groups (top) and clusters (bottom). 
The inner and outer contours in the plots correspond to $68\%$ and $95\%$ confidence levels of the posterior distributions for SIDM and CDM halos with and without AC. 
For the groups, we find that overall there is no preferred range for $\beta$. 
For the clusters, on the other hand, our analysis finds that negative values of $\beta$ are preferred for CDM and CDM+AC~halos, in contrast to Ref.~\cite{Schaller:2014gwa}, when $\beta$ is allowed to float freely. 
However, as in the Newman et al.\ analyses, we have assumed $\beta$ is constant with radius.
In Sec.~\ref{sec:numericalresults_sims}, we revisit $\beta$ in the context of SIDM simulations and show that, despite neglecting its radial dependence, our Jeans analysis nonetheless yields robust determinations for the cross section. 

%
\subsection{Mass-concentration relation \label{sec:M200c200}}

\begin{figure}[!t]
\centering
\includegraphics[width=0.49\textwidth]{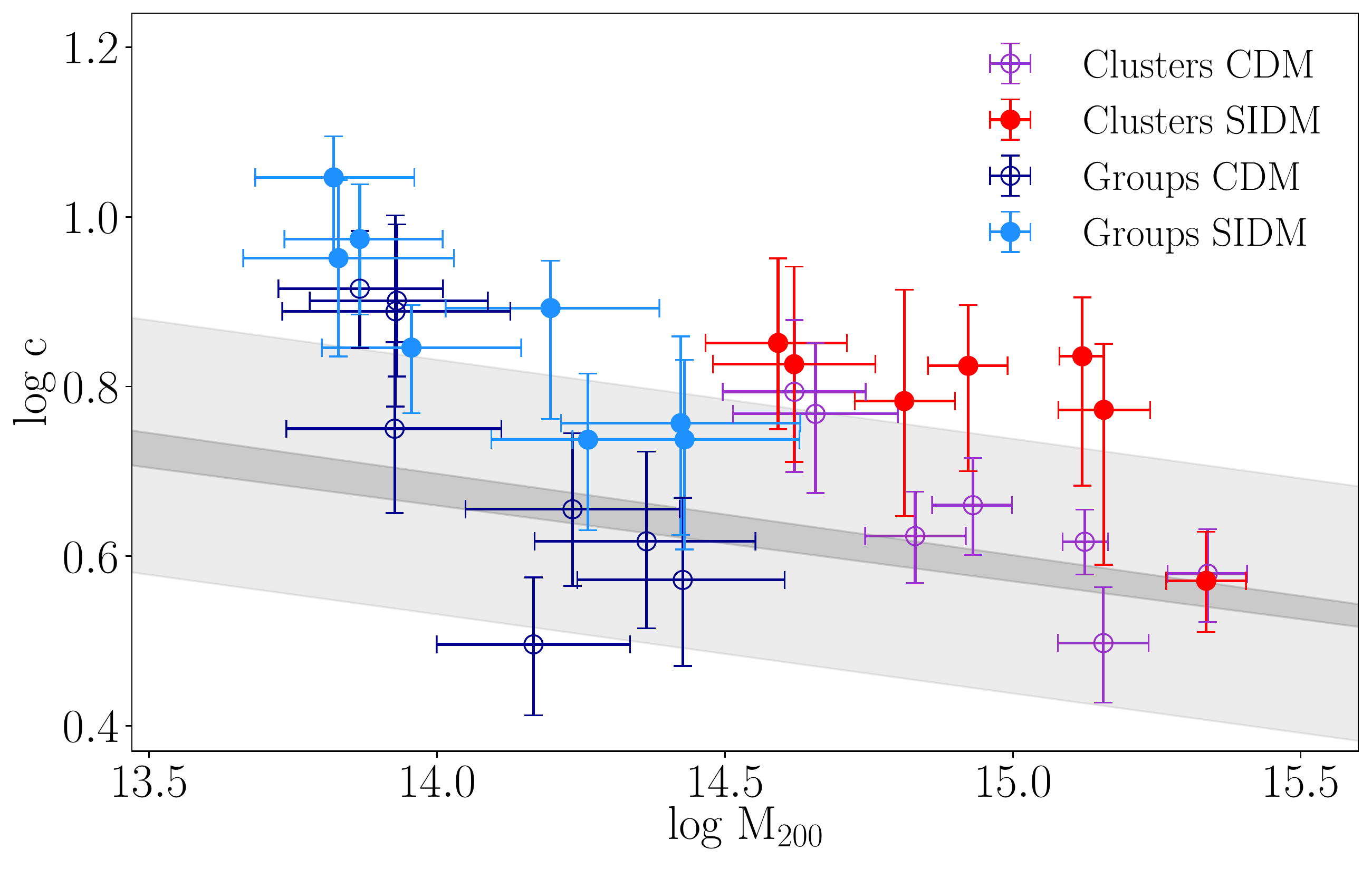}
\caption{\it $M_{200}$ versus $c$ for galaxy groups and galaxy clusters from our MCMC fits for SIDM and CDM (without AC). 
The darker band shows the median MCR from CDM-only simulations for the redshift range $0.19 < z < 0.45$ for these systems. 
The lighter band represents the uncertainty of $\pm 0.15\,\mathrm{dex}$ assumed in our MCMC prior (for a median value of $z = 0.3$).}
\label{fig:MCR_plots_obs}
\end{figure}

For SIDM, since the outer halo beyond $r_1$ has a negligible collision rate, the profile is expected to coincide with the corresponding CDM profile the halo would have in the absence of collisions.
We therefore expect the values of $M_{200}$ and $\conc$ obtained by matching in the Jeans model to represent cosmologically reasonable values obeying the MCR from CDM simulations.
Fig.~\ref{fig:MCR_plots_obs} shows our results for both groups and clusters, comparing our MCMC fits for SIDM (closed points) and CDM (open points) without AC.
The darker band shows the median MCR from Ref.~\cite{Dutton:2014xda} for the range of redshifts for the observed systems, and the lighter band shows the uncertainty $\pm 0.15$\,dex assumed in our prior.
Despite some scatter, our SIDM results are systematically shifted toward higher concentrations, while our CDM fits are in better agreement with the median MCR.

Intuitively, it makes sense that there is a trade-off between concentration and cross section if the central density is fixed by observations.
Larger $c$ increases the central density, while collisions typically reduce it.
The fact that our SIDM fits are shifted toward higher $c$ than predicted from the MCR may indicate a bias in the Jeans model.
On the other hand, since the group and cluster observations are all strong lenses, selection bias may play a role in the observations, which is known to return higher concentrations due to the halo orientations along the line of sight~\cite{Sereno:2014uqa}. 
It is therefore not obvious that our fits should lie along the median MCR.
Another consideration is that the MCR is a CDM-based prediction and may not be reflected in the outer regions of SIDM halos as we have in mind.

\section{Comparison to simulations \label{sec:sims}} 

In this section, we confront our methods against hydrodynamical simulations by Robertson et al.~\cite{Robertson:2017mgj}.
Re-simulating two clusters from the Cluster-EAGLE (CE) project~\cite{2017MNRAS.471.1088B,2017MNRAS.470.4186B} with $\sigma/m = 1\,  \cmg$, these were the first cosmological simulations on cluster scales including both baryonic physics and self-interactions.
The two clusters, dubbed CE-05 and CE-12, have virial masses $M_{\rm 200} = 1.4 \times 10^{14}\,\Msun$ and $3.9 \times 10^{14}\,\Msun$ at $z=0$, respectively, comparable to the sample of groups.\footnote{In comparison, the groups span $z=0.21$--$0.45$.}
Here we construct a set of mock observables from these simulations, analogous to those fit for the Newman et al.~\cite{Newman:2015kzv} groups.
Each set of observables is generated for a random line of sight and includes only particles located within $5\,r_{200}$ of the center of the halo (centered on the most bound particle)~\cite{Robertson:2017mgj}.
Then, we fit these observables using our Jeans approach to extract $\sigma/m$.
Encouragingly, we obtain values of $\sigma/m$ in the range $1$--$2 \, \cmg$, depending on various systematic assumptions, which we discuss below.

\begin{figure}[!t]
\centering
\includegraphics[width=0.49\textwidth]{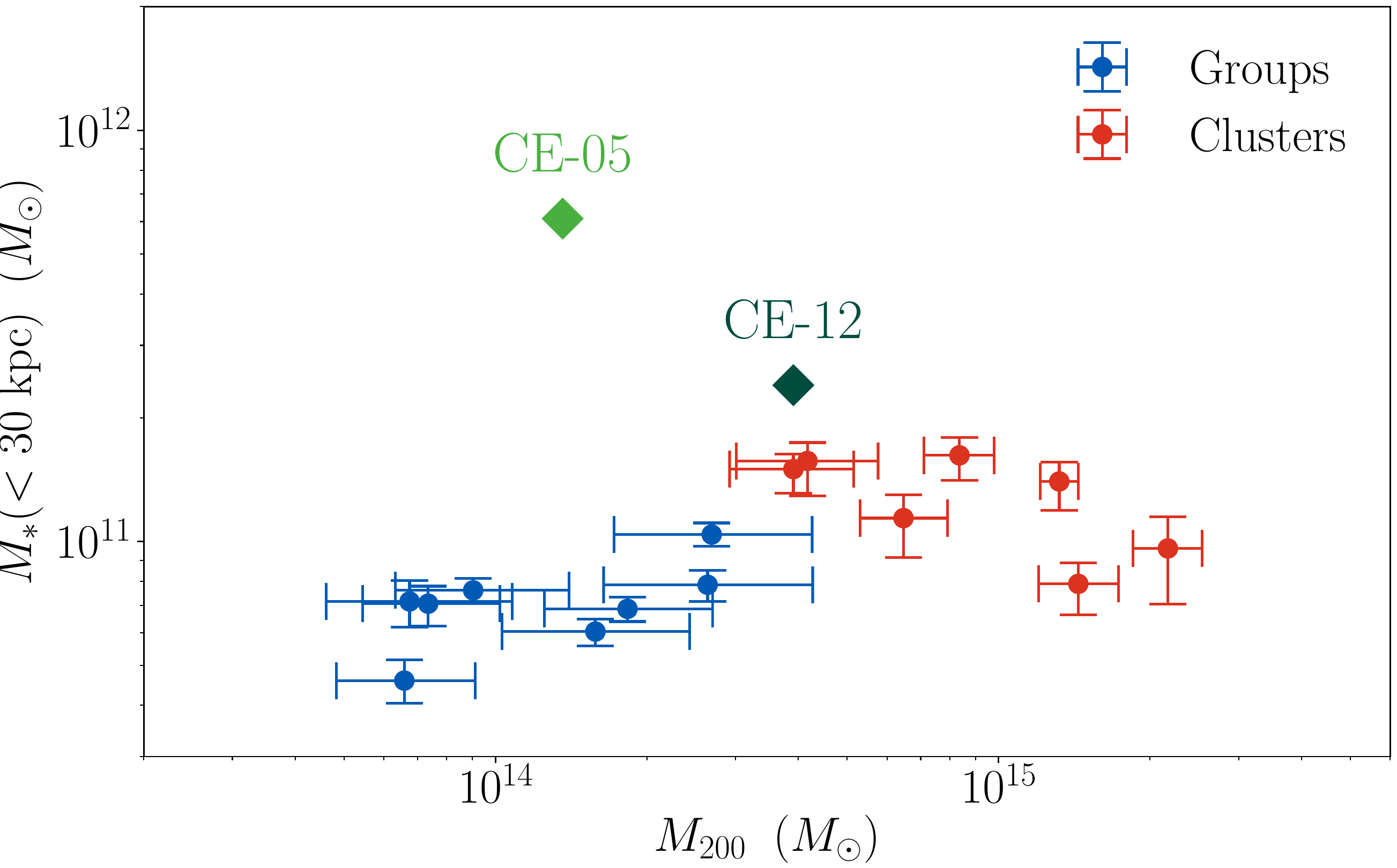}
\includegraphics[width=0.49\textwidth]{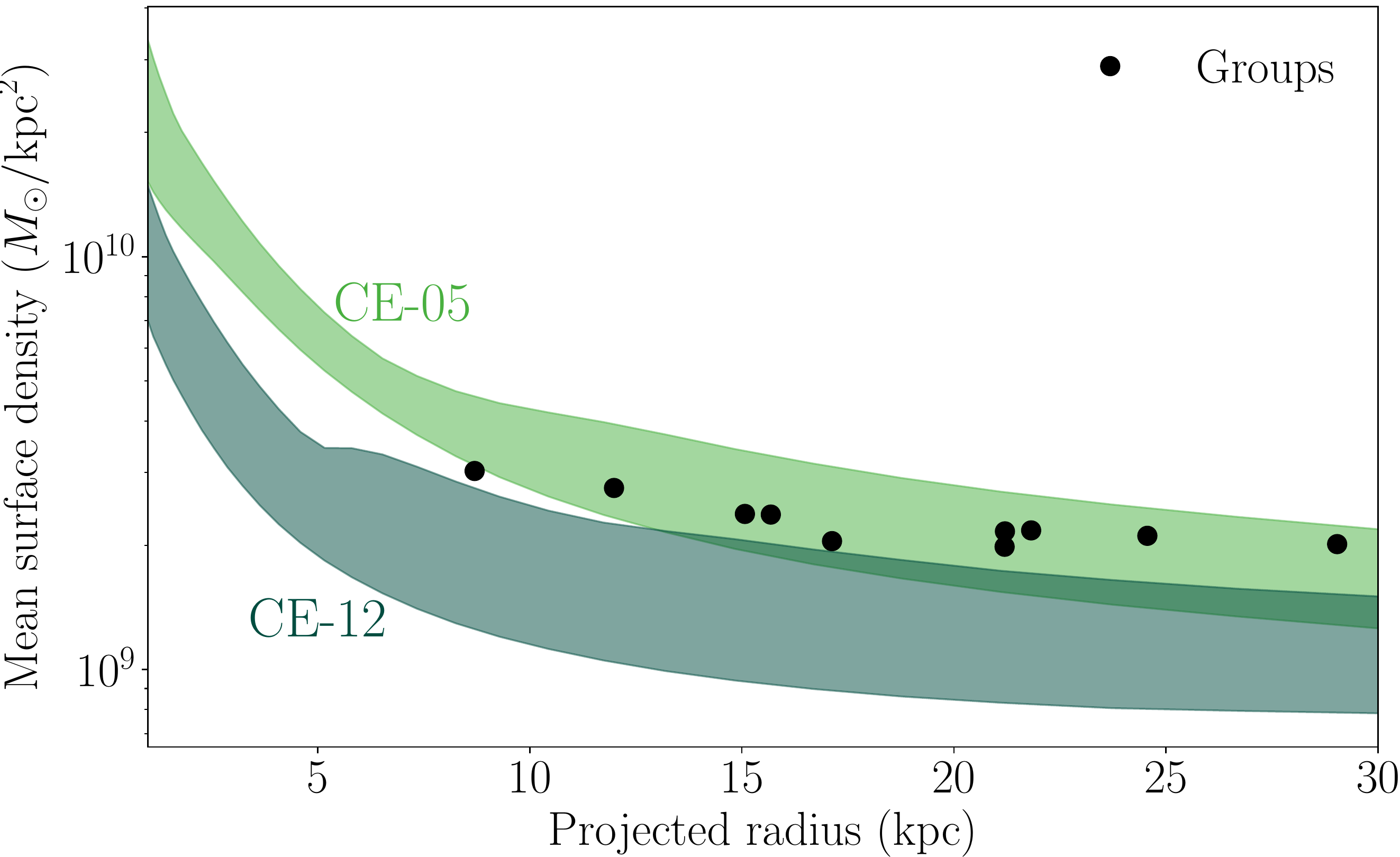}
\caption{\it Left: Central galaxies in CE simulations with SIDM (diamonds) have larger stellar masses $M_\star$ enclosed within $30 \, \mathrm{kpc}$, for given virial mass $M_{200}$, compared to values obtained in our fits for observed groups (blue points) and clusters (red points).
Right: For ten group lenses, points show their critical surface densities and effective Einstein radii (by azimuthally averaging their convergence profiles)~\protect\cite{Newman:2015kzv}.
Shaded bands show range of mean enclosed surface densities for CE-05 and CE-12 for 1000 random lines of sight.}
\label{fig:sim_mass}
\end{figure}

The comparison between groups and CE simulations is not strictly apples-to-apples and several comments are in order.
First, CE simulations for CDM produce central galaxies that are systematically more massive compared to observations~\cite{2017MNRAS.470.4186B}.
Similarly for the SIDM runs~\cite{Robertson:2017mgj}, BCG stellar masses are significantly larger compared to the groups and clusters, shown in Fig.~\ref{fig:sim_mass} (left).
Additionally, the two CE clusters have quite distinct inner halo profiles from one another~\cite{Robertson:2017mgj}.
Spherically-averaged, CE-12 has a flat profile with a large $100\, \mathrm{kpc}$ core, while CE-05 has no discernible core and its inner profile remains steep, scaling as $\rho_{\rm dm} \sim r^{-0.6}$.
While this diversity may be an artifact of the gravitational influence of BCGs that are too large, the CE halos are consistent with (bracketing at opposite extremes) the range of halos seen in larger volume SIDM simulations with more realistic-sized BCGs~\cite{Robertson:2018anx}.

Second, there is a selection bias on the groups given that they are all strong lenses. 
Fig.~\ref{fig:sim_mass} (right) shows the critical surface densities and effective Einstein radii for the groups~\cite{Newman:2015kzv} (points).
The shaded bands bracket the range of mean surface density profiles (enclosed within a given projected radius) for the CE clusters for different lines of sight.
While CE-05 seems comparable to the observed groups, CE-12 is less concentrated and it is unlikely for this system to yield a comparable strong lens to be included in the sample.

Lastly, the number of CE systems (two) is much smaller than the number of groups (eight) or clusters (seven) in our analysis.
Hence we inflate our sample by considering four lines of sight for each of CE-05 and CE-12, and treat each projection as an independent system for a total of eight systems, to have a comparably sized sample of mock observations.
Larger volume runs for SIDM with baryons have been performed within the BAHAMAS project, but the resolution is not yet sufficient for BCG stellar kinematics~\cite{Robertson:2018anx}.

In the remainder of this section, we first construct the same set of observables from the simulations as we have for the sample of groups~\cite{Newman:2015kzv}.
Then we perform an MCMC analysis to fit these systems based on the Jeans model.

\subsection{Mock observables \label{sec:MockObsSims}}

\begin{figure}[!t]
\includegraphics[width=0.32\linewidth]{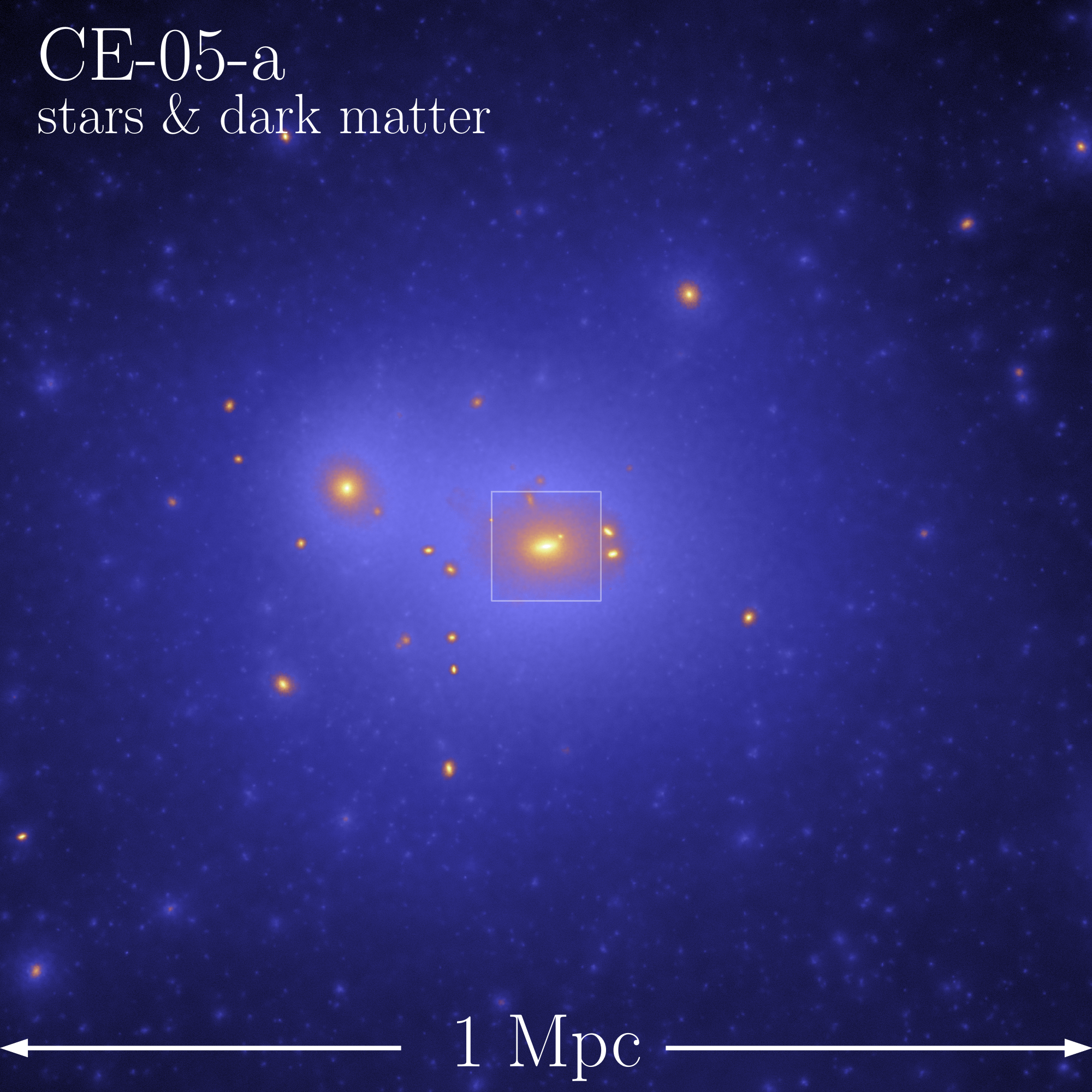}
\includegraphics[width=0.32\linewidth]{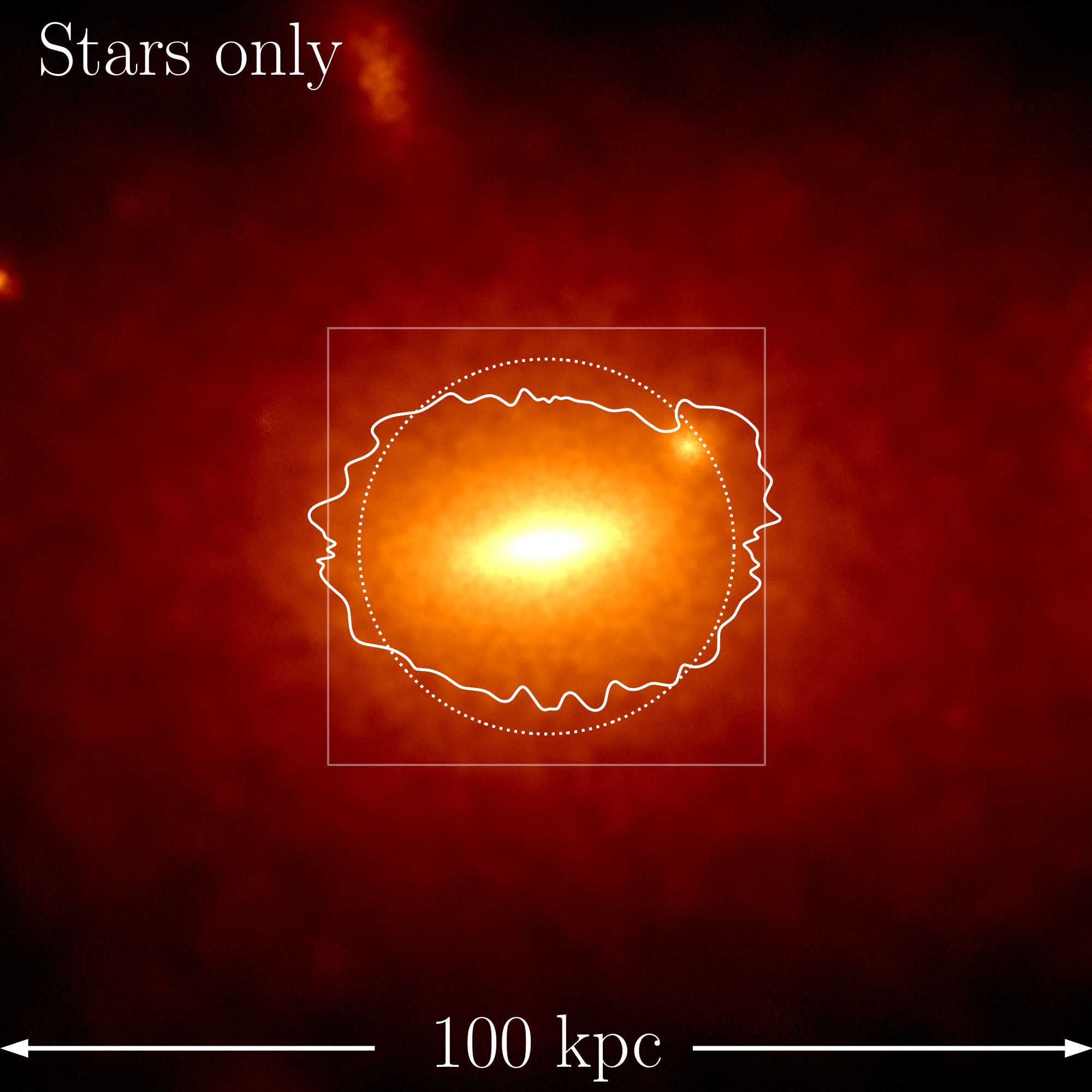}
\includegraphics[width=0.32\linewidth]{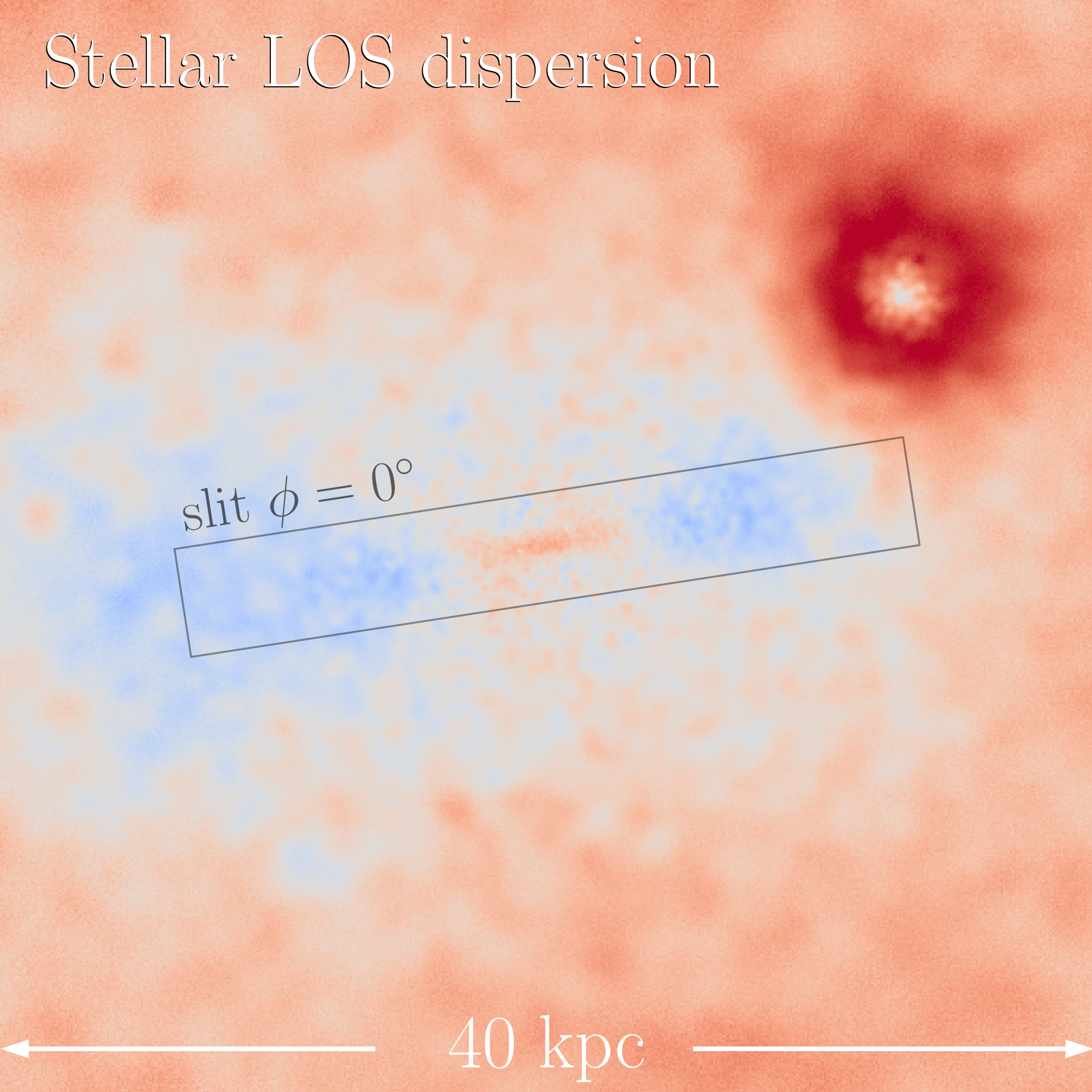}
\caption{\it Left: Dark matter plus stars. Center: Stars. Right: Stellar velocity dispersion. Squares denote inset region for the subsequent plot to the right.}
\label{fig:sim_panels}
\end{figure}

Our approach to mock observations is nicely summarized in Fig.~\ref{fig:sim_panels}. 
The left panel gives an overview of CE-05 on large scales, showing the projected mass densities for both dark matter (blue) and stars (orange-yellow).
The center panel shows a zoomed-in view of the central BCG.
This image is used to extract the stellar density profile.
With respect to strong lensing, the solid contour shows the tangential critical curve for this system, while the dashed circle shows the effective Einstein radius obtained by azimuthally averaging the projected mass density. 
The right panel is a further zoomed-in view of the BCG which shows the stellar line-of-sight velocity dispersion.
The red (blue) regions indicated hotter (colder) stellar kinematics with higher (lower) velocity dispersions.
The gray rectangle indicates the geometry of the slit, here aligned along the major axis of the BCG.
The slit is subdivided into nine bins along its length, which is used to compute the binned line-of-sight dispersions.
We describe these observables in more detail below.

{\it Stellar density profile:} To determine the stellar profile, we adopt an observationally-driven approach, rather than fitting the 3D profile directly.
First, we create a mock image of the BCG by projecting the stellar mass density along the line of sight, as shown in Fig.~\ref{fig:sim_panels} (center) for CE-05.
This image covers $(100\,\mathrm{kpc})^2$ in area and is divided into $1024\times 1024$ pixels.
To avoid issues of granularity, each star particle has been smeared with a 3D Gaussian whose width is set by the distance to its 8th nearest neighbor.

Next, we fit the BCG image with an ansatz for the surface density profile
\be \label{eq:Moffat}
\Sigma_\star(R) = \frac{\Sigma_0}{(1 + R^2/a_\star^2)^n} \, .
\ee
This function is known as a Moffat profile and is used to model seeing~\cite{1969A&A.....3..455M}.
It also happens to provide a good fit to the BCGs, though this is not motivated on physical grounds.
We allow for a BCG-halo offset and ellipticity in our fit by writing $R$ as
\be
R = \sqrt{ A(x-x_0)^2 + 2B(x-x_0)(y-y_0) + C(y-y_0)^2 },
\ee
where
\be
A = q \cos^2\theta + q^{-1} \sin^2 \theta \, , \quad
B = (q - 1/q) \sin\theta \cos\theta  \, , \quad
C = q \sin^2\theta + q^{-1} \cos^2 \theta \, .
\ee
The seven parameters of each fit are $\Sigma_0$, $a_\star$, $n$, the minor-to-major axis ratio $q$, position angle $\theta$, and offset $x_0$, $y_0$.
We determine their best-fit values, which are given in Table~\ref{tab:stellar_fits}, by fitting to pixels within $10\,\mathrm{kpc}$ of the halo center and assuming a fixed fractional uncertainty.
BCG-halo offsets appear to be negligible and hereafter we take the BCG and halo centers to coincide.

Overall, the BCGs have a sizable ellipticity.
Since our Jeans analysis is limited to spherical symmetry, we circularize\footnote{Circularization maps iso-density ellipses to circles and amounts to the replacement $\Sigma_\star(x,y) \to \Sigma_\star(R)$ for fixed $\Sigma_0$, $a$, $n$.} and de-project the surface density $\Sigma_\star(R)$ assuming spherical symmetry to obtain the 3D stellar density.
For the Moffat profile, de-projection can be done analytically using an inverse Abel transform to obtain
\be \label{eq:rho_Moffat_deproj}
\rho_\star(r) = \frac{\Sigma_0 \Gamma(n+\tfrac{1}{2})}{\sqrt{\pi} a_\star \Gamma(n) (1+r^2/a_\star^2)^{n+1/2}} \, ,
\ee
where $\Gamma$ is the gamma function.
Eq.~\eqref{eq:rho_Moffat_deproj} provides a good fit to the true spherically-averaged 3D profile obtained directly from the simulations and the resulting parameters are also given in Table~\ref{tab:stellar_fits} (denoted ``sph.~avg.'').
A comparison of the numbers for $(\Sigma_0, a_\star, n)$ in Table~\ref{tab:stellar_fits} shows how well our BCG image de-projection reproduces the true spherically-averaged 3D stellar profile.
For CE-12 there is good agreement across different lines of sight, while for CE-05 there is a larger scatter, signaling a breakdown of spherical symmetry.\footnote{The biggest outlier, CE-05-b, has a line of sight that happens to be oriented along the BCG minor axis.
This causes the stellar density to be comparatively more spread out in projected radius.}

\begin{table}[!b]
\centering
\begin{tabular}{|l|c|c|c|c|c|c|c|}
\hline
Name & 
$\Sigma_{0} \; (\Msun/{\rm kpc}^2)$ & 
$a_\star$ (kpc) &
$n$ &
$q$ &
$R_0$ (kpc) &
$R_{\rm Ein}$ (kpc)&
$M_{200} \; (\Msun)$
\\
\hline
CE-05-a 
& $2.81 \times 10^{10}$ & 1.26 & 1.12 & 0.46 & 0.13 & 17.2 & $1.16 \times 10^{14}$\\
CE-05-b 
& $1.22 \times 10^{10}$ & 2.64 & 1.41 & 0.82 & 0.10 & 15.1 & $1.80 \times 10^{14}$\\
CE-05-c 
& $2.73 \times 10^{10}$ & 1.31 & 1.13 & 0.50 & 0.14 & 18.2 & $1.43 \times 10^{14}$\\
CE-05-d 
& $1.84 \times 10^{10}$ & 1.92 & 1.31 & 0.60 & 0.10 & 15.2 & $2.20 \times 10^{14}$\\
\hline
\hline
CE-05 (sph.~avg.)& $2.10 \times 10^{10}$ & 1.47 & 1.11 & - & - & - & $1.36 \times 10^{14}$\\
\hline
\hline
CE-12-a 
& $7.96 \times 10^{9}$  & 0.85 & 0.80 & 0.49 & 0.08 & 6.3 & $1.62 \times 10^{14}$\\
CE-12-b 
& $9.25 \times 10^{9}$  & 0.77 & 0.80 & 0.55 & 0.10 & 6.9 & $2.06 \times 10^{14}$\\
CE-12-c 
& $1.03 \times 10^{10}$ & 0.75 & 0.81 & 0.63 & 0.09 & 8.3 & $3.42 \times 10^{14}$\\
CE-12-d 
& $9.14 \times 10^{9}$  & 0.74 & 0.78 & 0.56 & 0.09 & 6.9 & $1.93 \times 10^{14}$\\
\hline
\hline
CE-12 (sph.~avg.)& $8.47 \times 10^{9}$ & 0.86 & 0.83 & - & - & - & $3.91 \times 10^{14}$\\
\hline
\hline

\end{tabular}
\caption{\it Stellar profile parameters, effective Einstein radius, and virial mass obtained for four lines of sight (a-d) for each simulated cluster CE-05 and CE-12.
Projected BCG offset from halo center is $R_0 = \sqrt{x_0^2 + y_0^2}$. 
For comparison, parameters for spherically-averaged baryon profile fit directly from simulations (``sph.~avg.''), as well as the true $M_{200}$ value, are also provided.}
\label{tab:stellar_fits}
\end{table}

{\it Stellar velocity dispersions:} We compute the mass-weighted line-of-sight velocity dispersion of star particles from the simulations.
For example, Fig.~\ref{fig:sim_panels} (right) shows the 2D velocity dispersion profile for CE-05 along one line of sight.
Star particles have been smeared relative to their nearest neighbors, as discussed above.
It is clear that the velocity dispersion of CE-05 is anisotropic: stars are kinematically colder (blue) along the major axis, compared to hotter regions (red) off-axis.\footnote{The hot ring in this figure is a substructure, which by itself is dynamically cold but appears hot when averaged with the BCG in projection.}
Observationally, the velocity dispersion is measured only within a roughly one-dimensional region corresponding to the slit (gray rectangle).
Given the anisotropy, we are wary that choosing a fixed slit orientation may introduce a systematic bias.
To assess this, we consider three different orientations for our slit, described by a relative angle $\phi=0^\circ$, $45^\circ$, and $90^\circ$ with respect to the BCG major axis.
We create three parallel sets of mock observables, one for each value of $\phi$.

Our observables are line-of-sight velocity dispersions averaged within 2D spatial bins corresponding to a slit geometry.
For the slit, we take a rectangle of physical length $\ell = 27\,\mathrm{kpc}$ and width $w=4\,\mathrm{kpc}$, centered on the halo.
The slit is binned along its length into 9 equal-sized bins ($3 \times 4 \, \mathrm{kpc}^2$) to yield the line-of-sight dispersion profile. 
We additionally smear the locations of star particles with a Gaussian point spread function (PSF) of width $3\,\mathrm{kpc}$ to mimic the effect of seeing.

\begin{figure}[!t]
  \begin{center}
    \includegraphics[width=0.98\textwidth]{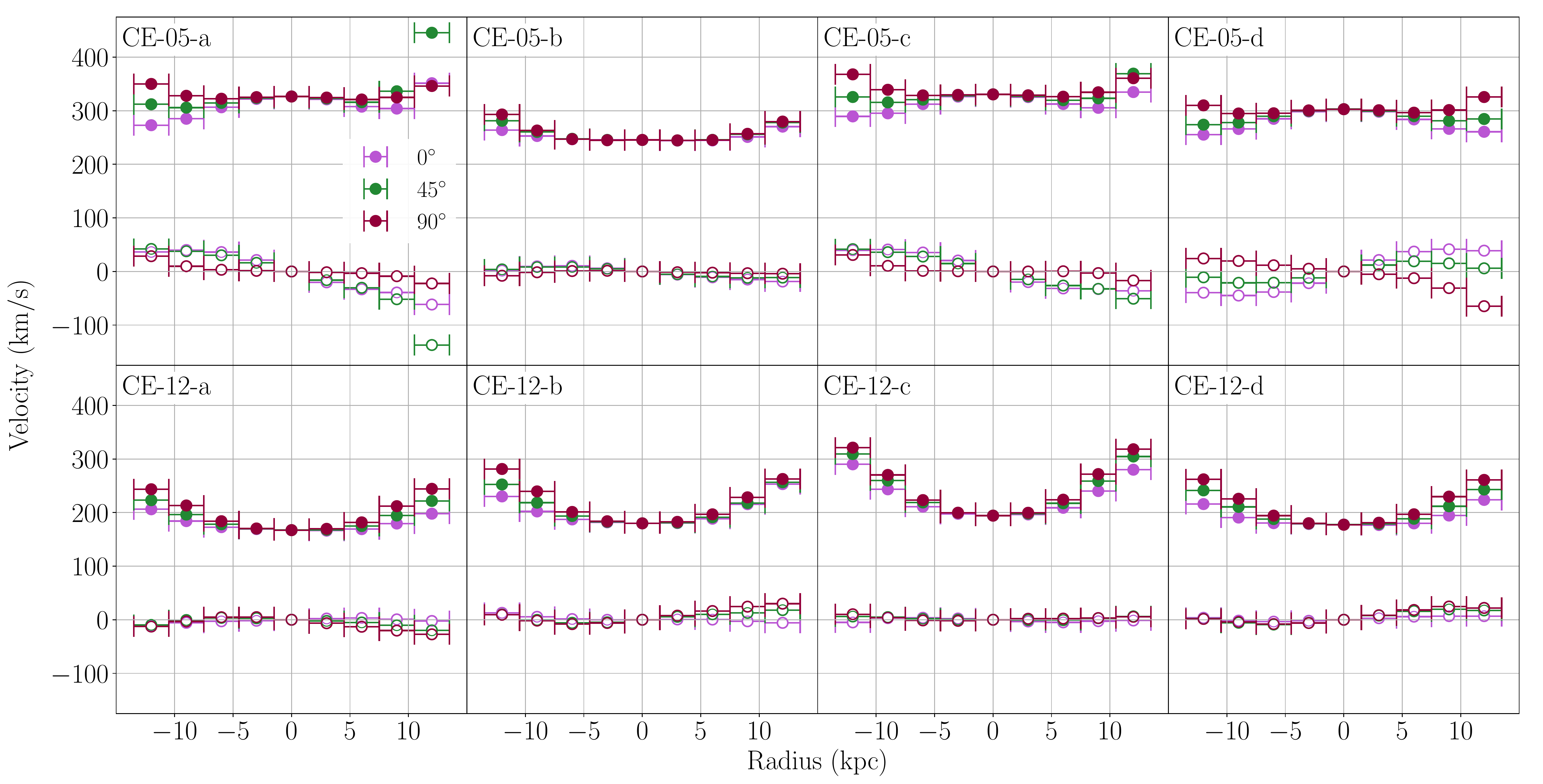}
  \end{center}
  \caption{\it Binned stellar line-of-sight velocity dispersion (solid points) and relative line-of-sight velocity (open points) for simulated clusters CE-05 and CE-12, along four lines of sight (a-d). 
Slit length is oriented with relative angles $0^\circ$, $45^\circ$, and $90^\circ$ relative to the BCG major axis.
}
  \label{fig:losvel_bins}
\end{figure}

Binned line-of-sight velocity dispersions are shown in Fig.~\ref{fig:losvel_bins} (solid points).
The dispersion profiles for CE-05 can be either increasing or decreasing with radius, depending on both line of sight and slit angle $\phi$.
The profiles for CE-12, by contrast, are uniformly increasing independent of line of sight or position angle.
We also show the binned line-of-sight velocity (open points) relative to the central bin which is normalized to zero.
This confirms our expectation that the BCGs are predominantly dispersion-supported with little rotational motion.

{\it Gravitational lensing:} We determine the effective Einstein radius from the azimuthally-averaged total mass projected along the line of sight. 
For the critical densities, we assume that the lens and source distances are such that $\Sigma_{\rm crit} = 10^{9.3} \, \Msun/\mathrm{kpc}^{2}$ for CE-05 and $10^{9.2} \, \Msun/\mathrm{kpc}^{2}$ for CE-12.
The corresponding Einstein radii, given in Table \ref{tab:stellar_fits}, are around $6$--$18\,\mathrm{kpc}$, which is comparable to the groups sample.
In our fits, we require the mean convergence within the Einstein radius to be unity with a $5\%$ uncertainty, following Eq.~\eqref{eq:lensing_chisq}.

It is worthwhile mentioning again that CE-12 is unlikely to be part of a strong lensing sample.
Our $\Sigma_{\rm crit}$ value for CE-12 is outside the range of critical densities $10^{9.3} $--$10^{9.5} \, \Msun/\mathrm{kpc}^{2}$ spanned by the group sample.
This lower value could be achieved for lens and source redshifts around $z_L \sim 0.8$ and $z_S \sim 7$, respectively, which would be on the extreme side of cluster lensing surveys~\cite{Postman:2011hg}.
Alternatively, one might envision some external convergence that has been subtracted out.

We can assess from the simulations whether the azimuthally-averaged Einstein radius bears any resemblance to the actual tangential critical curves.
First, we compute $1\,\mathrm{Mpc}^2$ images for the projected densities of dark matter, stars, gas, and black holes, each divided into $2048 \times 2048$ pixels.
For stars and dark matter, particles are smeared with a Gaussian whose width is set by the distance to their 8th nearest neighbor.
One example is shown in Fig.~\ref{fig:sim_panels} (left) for CE-05.
Gas particles are also smeared but with a width set by their respective smoothing lengths.\footnote{Smoothing lengths are used in the smoothed-particle hydrodynamics calculations in the Cluster-EAGLE simulations. Each gas particle is assigned a smoothing length at each time-step of the simulation, which is approximately the distance to the 58th nearest gas particle.}
From here, it is straightforward to compute the total convergence $\kappa_{ij}$, where $i,j$ labels a given pixel.
Next, we compute the shear components as a discretized sum over pixels
\bea
\gamma^{(1)}_{ij} &=& - \frac{1}{\pi} \sum_{k,l} \kappa_{kl} \frac{(i-k)^2 - (j-l)^2}{(i-k)^2 + (j-k)^2 + \epsilon}, \\
\gamma^{(2)}_{ij} &=& - \frac{1}{\pi} \sum_{k,l} \kappa_{kl} \frac{2 (i-k) (j-l)}{(i-k)^2 + (j-k)^2 + \epsilon} \, ,
\eea
where $i,j,k,l$ are simply integers labeling the pixel number and $\epsilon$ is a small nonzero number introduced to regulate the term in the sum where $i=k$ and $j=l$.
Lastly, we use spline interpolation to determine the tangential critical curve from our discretized quantities.
Fig.~\ref{fig:sim_panels} (center) shows our result for the true critical curve (solid contour), compared with our idealized critical curve from the azimuthally-averaged Einstein radius (dashed circle).
Using the shoelace formula, we find that the area (and therefore the total enclosed mass) within both contours agree to better than $1\%$.

{\it Virial mass:} When analyzing the simulated halos, we use a Gaussian prior on the logarithm of $M_{200}$, with a standard deviation of $0.2$\,dex. For the central value of this Gaussian prior, we estimate $M_{200}$ using the kinematics of cluster member galaxies, similar to the group observations~\cite{Newman:2015kzv}.
However, since the values of $M_{200}$ obtained for CE-12 are systematically lower than their true values (shown below), we also repeat our analysis simply using the true $M_{200}$ as the central value of the prior for each halo.

To infer $M_{200}$ from the kinematics of cluster members, we use a scaling relation by Munari et al.~\cite{Munari:2013mh}, 
\begin{equation} \label{eq:M200_Munari}
\log \left(h(z) \, M_{200} / \Msun\right) = 13.98+ 2.75 \log\left( \frac{\sigma_{\rm LOS}}{500\,\mathrm{\kms}}\right)\, ,    
\end{equation}
where here $\sigma_{\rm LOS}$ is the line-of-sight velocity dispersion of galaxies in the cluster and $h(z)$ is the dimensionless Hubble parameter.
This relation was determined from CDM simulations and is the same scaling relation used by Ref.~\cite{Newman:2015kzv} to estimate $M_{200}$ for the observed groups.

For the simulated clusters, we calculate the line-of-sight velocity of each galaxy within $r_{200}$ of the cluster center and with a stellar mass $M_* > 3 \times 10^9\,\Msun$. 
To do this, we take all star particles identified as belonging to a particular galaxy by the SUBFIND algorithm~\cite{Springel:2000qu} and compute a mass-weighted mean velocity for these star particles. 
We then take the component of this velocity along the line of sight as the line-of-sight velocity of the corresponding galaxy. 
The velocity dispersion $\sigma_\mathrm{LOS}$ of the cluster is calculated as the standard deviation of the line-of-sight velocities of all cluster members above our stellar mass threshold. 

The main difference between this method and what was done observationally is that here we only use galaxies within $r_{200}$ of the cluster center, whereas Ref.~\cite{Newman:2012nv} included galaxies out to larger radii. 
Galaxies at larger radii are not virialized within the cluster potential, and so tend to have lower velocity dispersions than galaxies within the cluster. 
As such, we found that including these galaxies typically leads to $M_{200}$ estimates that are biased low with respect to the true values because the Munari et al.\ relation
was calculated using velocity dispersions for galaxies within $r_{200}$.

Note that there is a relatively large scatter in this relation (approximately $40\%$ scatter in $M_{200}$ at fixed $\sigma_\mathrm{LOS}$) so that the accuracy of the $M_{200}$ values obtained in this way is limited. This can be seen in Table~\ref{tab:stellar_fits} where we list the values for $M_{200}$ we get using the Munari et al.\ relation as well as the true values of $M_{200}$. 
For CE-05, the estimated values for $M_{200}$ for the different lines of sight are distributed around the true value, as expected. 
In contrast, for CE-12 they are systematically low, but still consistent with the true value of $M_{200}$ when accounting for the scatter in the Munari et al.\ relation.

To investigate the impact on our results, we run MCMCs using both the $M_{200}$ values derived from the galaxy member kinematics and the true value of $M_{200}$. 
We find that the Jeans model gives robust predictions for the cross section, independent of the exact values of $M_{200}$, discussed in the next section.

\subsection{Numerical results
\label{sec:numericalresults_sims}} 

The analysis for our mock observations follows the same Jeans-based approach as in Sec.~\ref{sec:numerics}.
Each mock data sample comprises observables (stellar kinematics and projected density, effective Einstein radius, and virial mass) for eight systems, corresponding to four lines of sight for each cluster, CE-05 and CE-12, which are each treated as independent systems.
Moreover, we consider three realizations of the data sample, each for a different slit angle for the stellar kinematics, $\phi = 0^\circ$, $45^\circ$, and $90^\circ$.
For the baryon density, we take $\rho_b(r) = \Upsilon_\star \rho_\star(r)$, where $\rho_\star$ is given in Eq.~\eqref{eq:rho_Moffat_deproj} with fixed stellar parameters given in Table~\ref{tab:stellar_fits}.
We allow for the overall normalization $\Upsilon_\star$ to float independently for each system, analogous to our group and cluster fits.
Adopting the same priors in Table~\ref{tab:mcmcparams}, the free parameters for each system are $M_{200}$, $\conc$, $\langle \sigma v \rangle/m$, $\Upsilon_\star$, and $\beta$.
We do not consider AC for the outer collisionless halo\footnote{For clusters simulated within the EAGLE project for CDM, dark matter halos are consistent with NFW profiles without significant AC~\cite{Schaller:2014gwa}.}, nor a gradient for $\Upsilon_\star$.

\begin{figure}[!t]
\centering
\includegraphics[width=0.49\textwidth,valign=t]{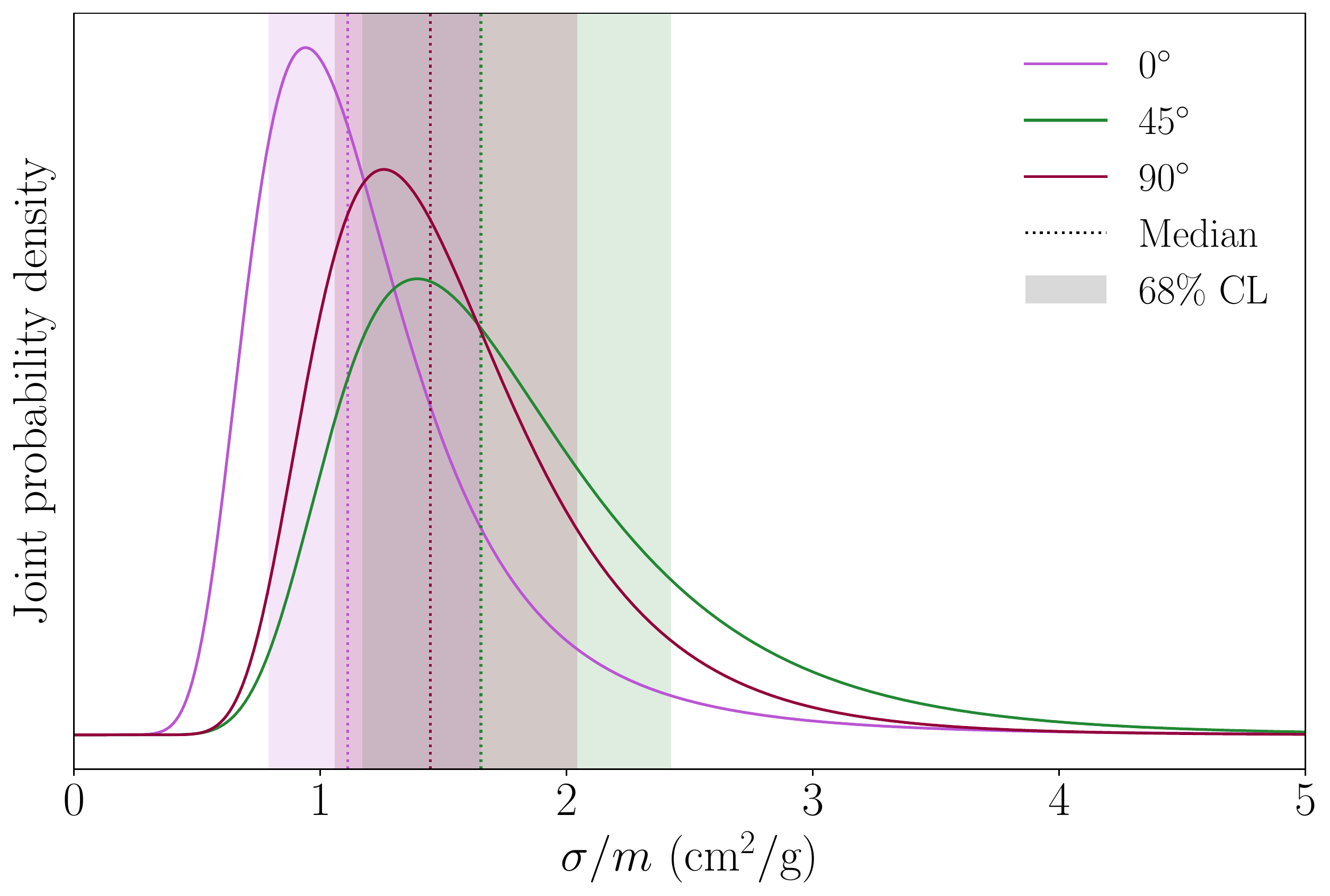}
\includegraphics[width=0.5\textwidth,valign=t]{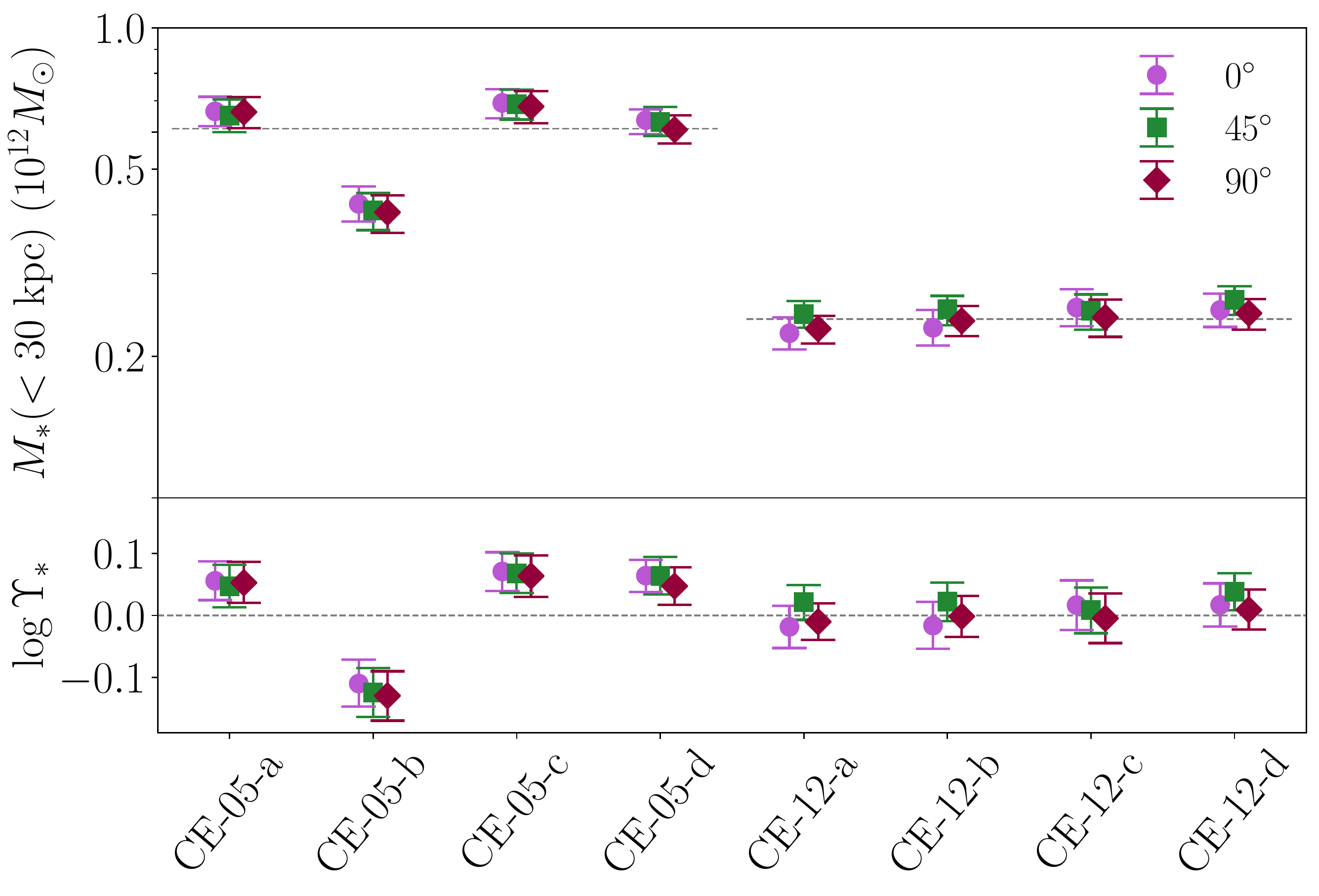}
\caption{\it Outputs from our mock dataset analysis for three different slit angles $0^\circ$, $45^\circ$ and $90^\circ$.
Each dataset is four lines of sight for each of CE-05 and CE-12.
Left plot: Joint probability density for the cross section. True value is $\sigma/m = 1\,\cmg$.
Right plot: Stellar mass within $30\,\mathrm{kpc}$ (upper panel) and mass-to-light ratio $\Upsilon_\star$ (lower panel). 
Dotted lines show the true values. 
}
\label{fig:sim_crosssec}
\end{figure}

Following Sec.~\ref{sec:numerics}, we perform an MCMC analysis for each data sample and construct the joint probability distributions for $\sigma/m$ from Eq.~\eqref{eq:jointprob}.
These curves are shown on the left plot in Fig.~\ref{fig:sim_crosssec}. 
The corresponding shaded bands indicate the $68\%$ CL intervals and the dotted lines are the median values.
These correspond to
\begin{equation}
\label{eq:sigmam_sim}
\sigma / m = 
\left\{ \begin{array}{ll}
1.1^{\,+0.5}_{\,-0.3}\,\cmg\; & \textnormal{slit angle $0^\circ$} \\[5pt]
1.7^{\,+0.8}_{\,-0.5}\,\cmg\; & \textnormal{slit angle $45^\circ$} \\[5pt]
1.4^{\,+0.6}_{\,-0.4}\,\cmg\; & \textnormal{slit angle $90^\circ$}
\end{array} \right. .
\end{equation} 
Our results are in mutual agreement for different choices of slit angle, albeit with a slight bias toward values of $\sigma/m$ larger than 1 $\cmg$.
This bias may be due to an underestimate of the age entering the rate equation \eqref{eq:rate}.
The mock observations are determined at redshift $z=0$, when the Universe is $20$--$50\%$ older than it was at the redshifts of the observed groups and clusters.
However, for consistency, we have fixed the mean age to be the same $t_0 = 5 \, {\rm Gyr}$.
There are indications from the simulations that the characteristic ages for CE-05 and CE-12 may be larger, e.g., the age since half of the BCG stellar mass was accumulated is around $8$--$9\,\mathrm{Gyr}$~\cite{Robertson:2017mgj}, which may be a rough proxy for the age of the system. 
Further study is required to improve our estimations for $t_0$.

Next, we verify other outputs from our analysis, starting with the BCG stellar densities in Fig.~\ref{fig:sim_crosssec} (right).
The upper panel shows the total stellar mass enclosed with $30\,\mathrm{kpc}$ inferred from our fits (points) compared to the true values (dotted lines).
These fitted values depend on our modeling and de-projection of the BCG surface densities, as well as our MCMCs converging on the true value $\Upsilon_\star$.
The bottom panel shows the latter compared to the true value $\Upsilon_\star = 1$.
The fact that our results successfully reproduce true stellar masses -- despite our very weak prior on $\Upsilon_\star$ -- provides a check that our Jeans analysis is able to determine both the baryon and dark matter densities simultaneously.
Moreover, there is little dependence on the different choice of slit angle, $\phi = 0^\circ$, $45^\circ$, or $90^\circ$.
The largest outlier is CE-05-b.
Because this line of sight is accidentally aligned with the BCG minor axis, one infers a reduced stellar surface density and a reduced stellar line-of-sight velocity dispersion (see Fig.~\ref{fig:losvel_bins}), which ultimately leads to a $\sim 30\%$ underestimate of the stellar mass.

\begin{figure}[!t]
\centering
\includegraphics[width=0.98\textwidth,valign=t]{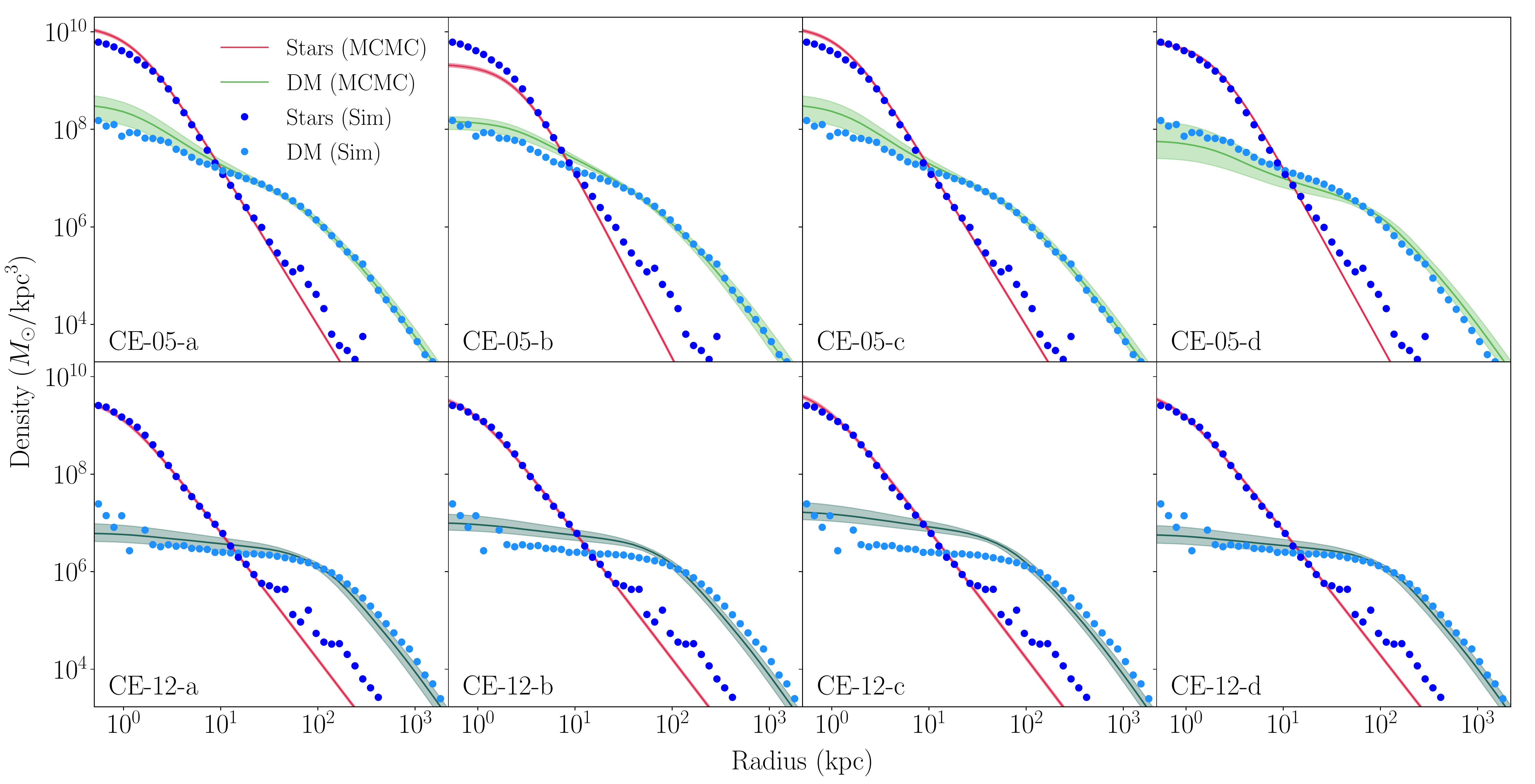}
\caption{\it Stellar and dark matter density profiles for CE-05 and CE-12 along four lines of sight (a-d) from our MCMCs ($68\%$ CL bands), plotted with the true stellar and dark matter density profiles from the simulation (dots).}
\label{fig:sim_dens_plots}
\end{figure}

In Fig.~\ref{fig:sim_dens_plots}, we show the dark matter and stellar density profiles from our Jeans analysis (solid bands) compared to the true spherically-averaged profiles from the simulations (dots).
The bands represent the $68\%$ scatter in profiles inferred from our MCMCs.
This is another check that our analysis is able to determine simultaneously the baryon and dark matter profiles, even in the inner region where baryons are dominant.
For the stars, our inferred profiles (red) are accurately determined and in good agreement with the true profiles, again with CE-05-b being the largest outlier.
(Note we do not model other stellar structures beyond the BCG that become apparent at radii $\gtrsim 50\,\mathrm{kpc}$.)
For the dark matter, our results are mostly consistent with the true profiles.
However, for CE-12, there is clear variation between lines of sight, with the largest outlier overestimating the dark matter central density by a factor of two.

\begin{figure}[!t]
\begin{center}
\includegraphics[width=0.65\textwidth]{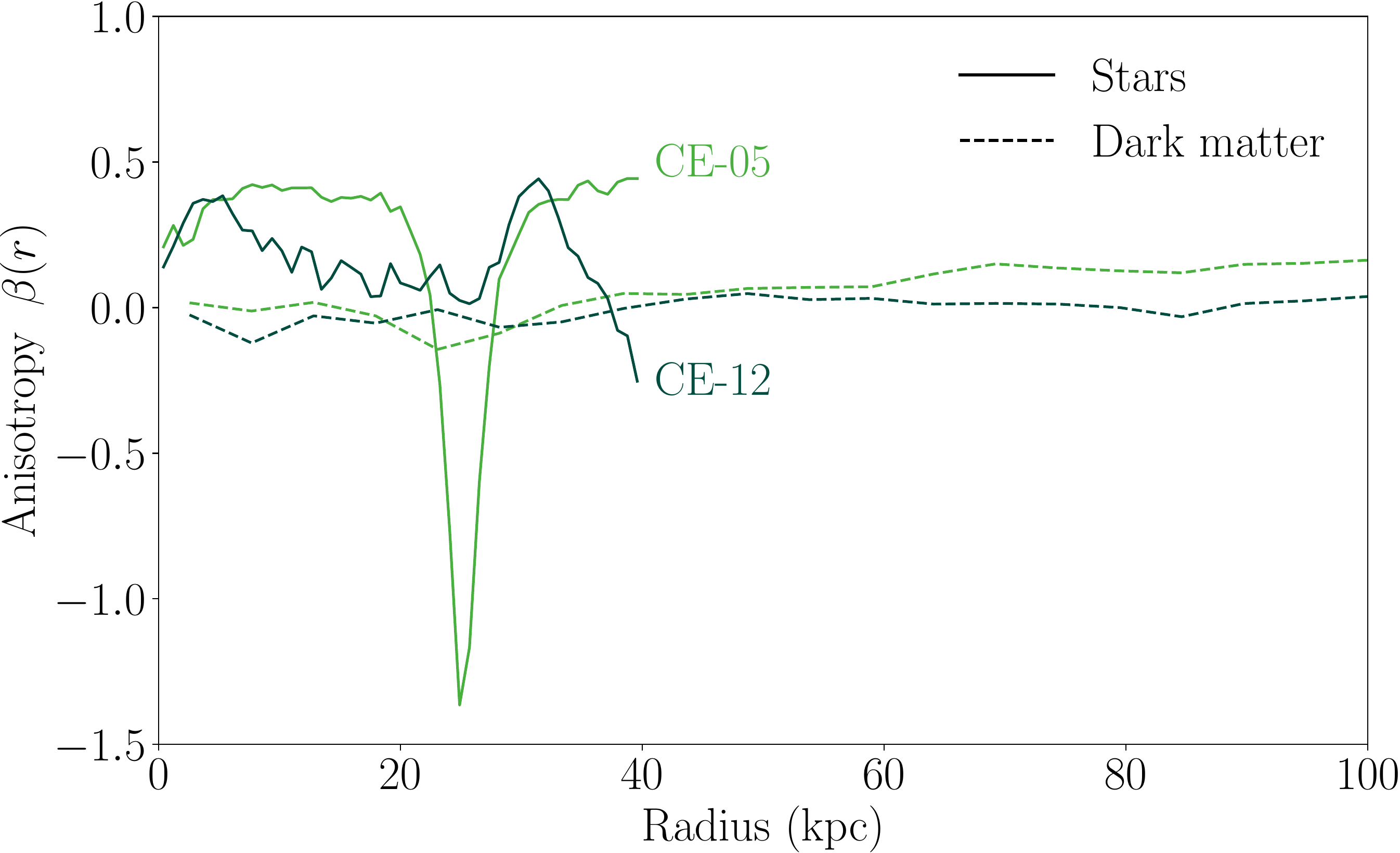}
\end{center}
\caption{\it Velocity dispersion anisotropy profile $\beta(r)$ from CE simulations for stars (solid) and dark matter (dashed).
}
\label{fig:n_body_stars}
\end{figure}

Next, we consider the stellar velocity dispersion anisotropy $\beta$. 
Fig.~\ref{fig:n_body_stars} shows the actual radial profile for $\beta$ within the central $40\,\mathrm{kpc}$ of the BCG (solid lines), with a preference for radially-biased orbits with $\beta > 0$.\footnote{Note that $\beta$ is not a frame invariant quantity. In Fig.~\ref{fig:n_body_stars}, we have computed $\beta$ in the rest frame of the BCG within $40\,\mathrm{kpc}$ for stars, and in the rest frame of the inner halo within $200\,\mathrm{kpc}$ for dark matter. In the latter case, distances are given with respect to the most bound dark matter particle, which is coincident with the center of the BCG at the $0.1\,\mathrm{kpc}$ level (cf.~Table~\ref{tab:stellar_fits}).}
While our analysis assumes $\beta$ is constant and within the range $|\beta| < 0.3$, it is clear that $\beta$ is neither constant nor wholly within the assumed range.
Nevertheless, we retain these assumptions to be consistent with our treatment of the observed groups and clusters.
Fig.~\ref{fig:n_body_stars} also shows the velocity anisotropy profiles for dark matter particles (dashed lines).
We find $\beta \approx 0$ for dark matter, as assumed in the Jeans model, since self-interactions isotropize velocities in the inner halo.\footnote{Ref.~\cite{Sokolenko:2018noz} has proposed an alternative SIDM Jeans framework extended to include $\beta \neq 0$ for dark matter.}

\begin{figure}[!t]
  \centering
  \includegraphics[width=0.98\textwidth]{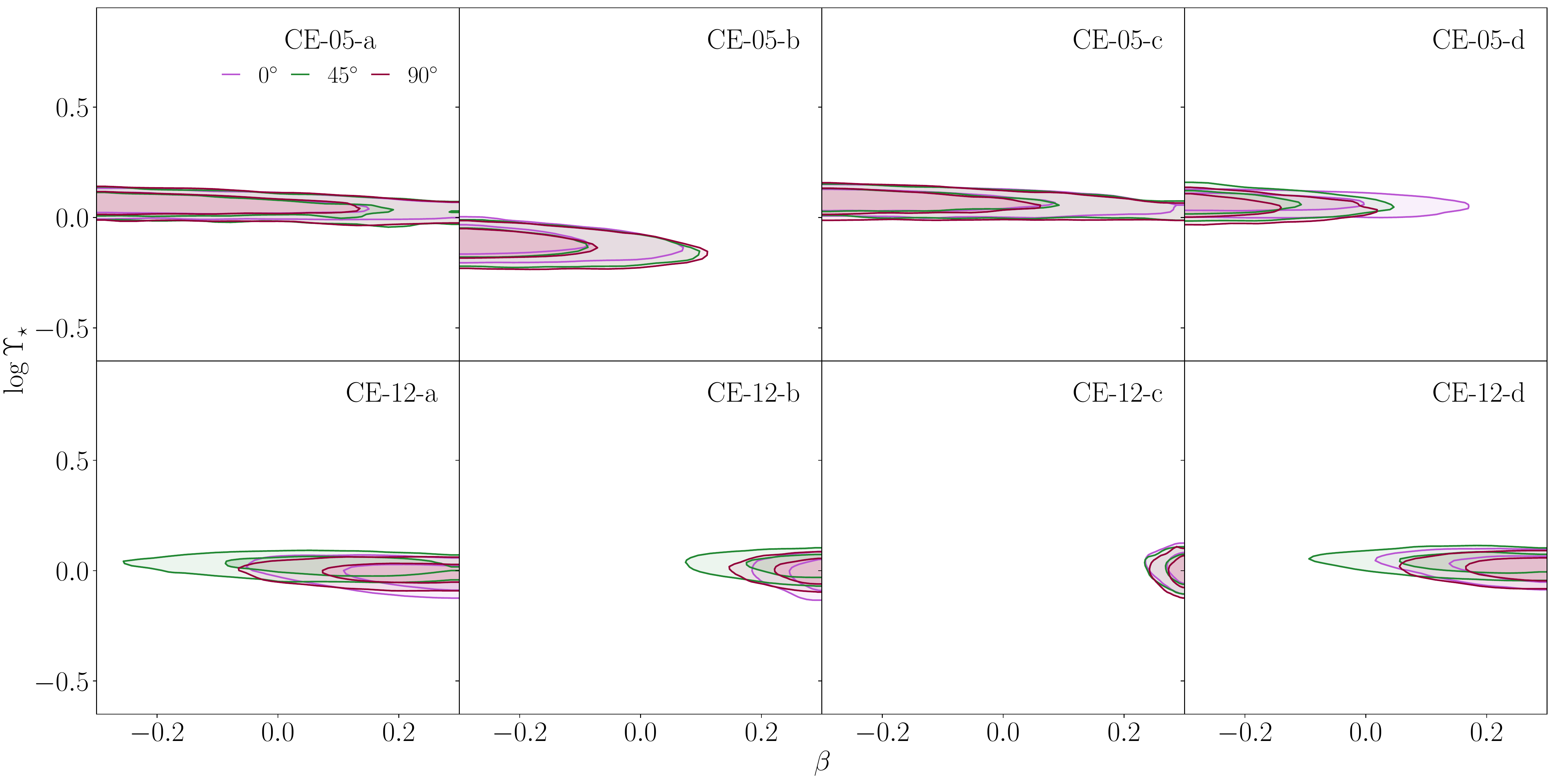}
  \caption{\it Mass-to-light ratio $\Upsilon_\star$ as a function of $\beta$ for the simulated clusters. The inner and outer contours in the plots indicate the $68\%$ and $95\%$ confidence levels of the posterior distributions from our analysis assuming SIDM halos without AC
  for slit angles of $0^\circ$, $45^\circ$ and $90^\circ$.}
  \label{fig:Y_vs_beta_sims}
\end{figure}

The preferred range for $\Upsilon_\star$ versus $\beta$ from our MCMCs is shown in Fig.~\ref{fig:Y_vs_beta_sims}.
Our fits are consistent with $\Upsilon_\star = 1$, as discussed above, and show no correlation between $\Upsilon_\star$ and $\beta$.
CE-12 prefers $\beta > 0$, as we expect, but interestingly CE-05 has a mild preference for tangential orbits with $\beta < 0$.
We have investigated further by repeating our MCMC analysis with different priors for $\beta$.
If we simply fix $\beta=0.3$ for all systems -- a rough proxy for the true $\beta$ profiles in Fig.~\ref{fig:n_body_stars} -- the resulting stellar and dark matter profiles, as well as the inferred $\sigma/m$, are largely unchanged from our quoted results.
On the other hand, if we relax our prior to allow for a larger range $|\beta| < 1$, we find that the inferred dark matter profiles for lines of sight CE-12-b,c yield better matches to the true profiles, however, at the expense of unrealistically large values $\beta \approx 0.4$--$0.6$.
Moreover, the inferred joint cross section is biased high and is in worse agreement with $1\,\cmg$.
All this is to say that our original prior $|\beta| < 0.3$ seems satisfactory for obtaining $\sigma/m$ reliably, despite the limitations we have raised.

\begin{figure}[!t]
\centering
\includegraphics[width=0.49\textwidth]{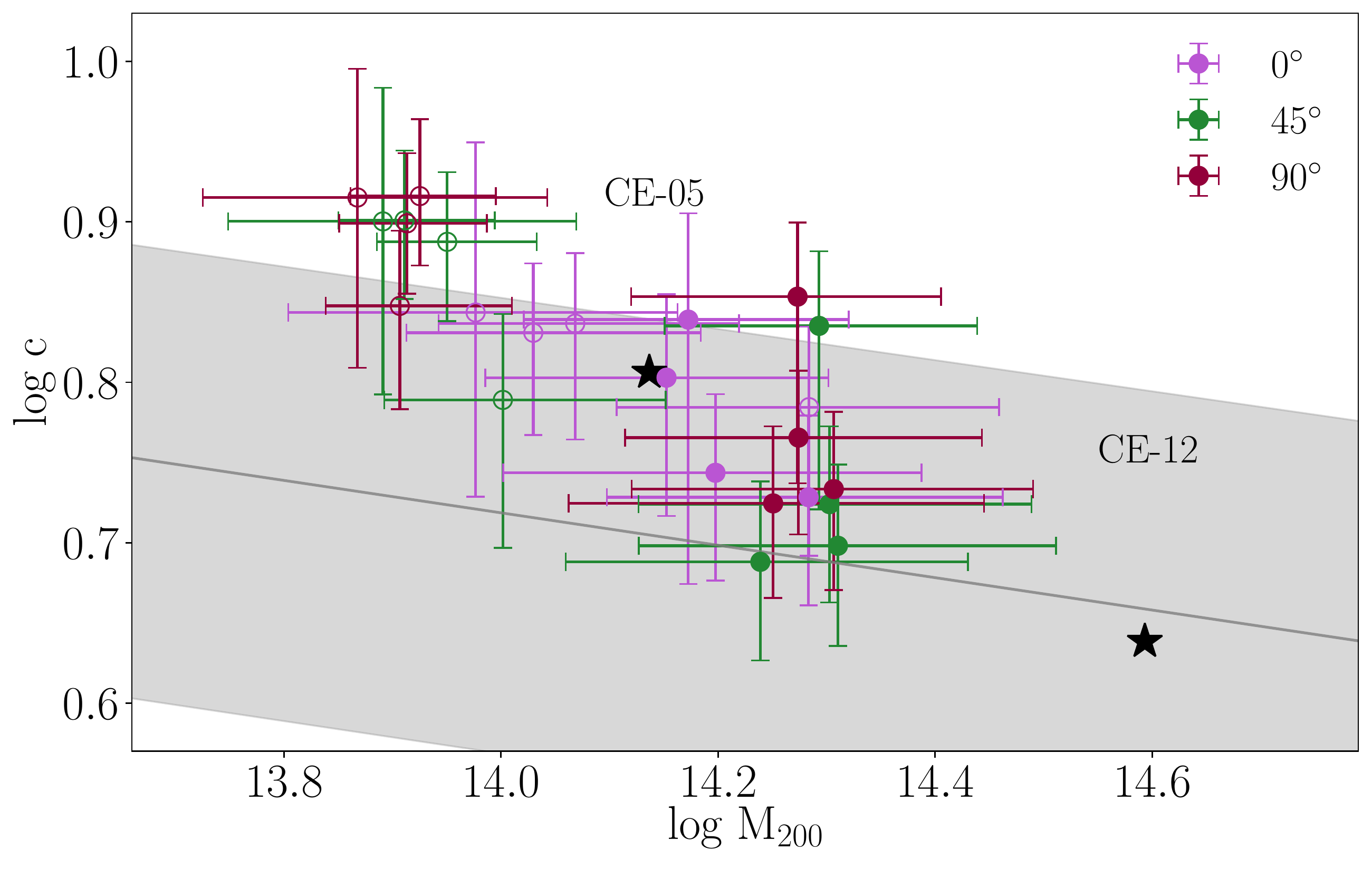}
\includegraphics[width=0.49\textwidth]{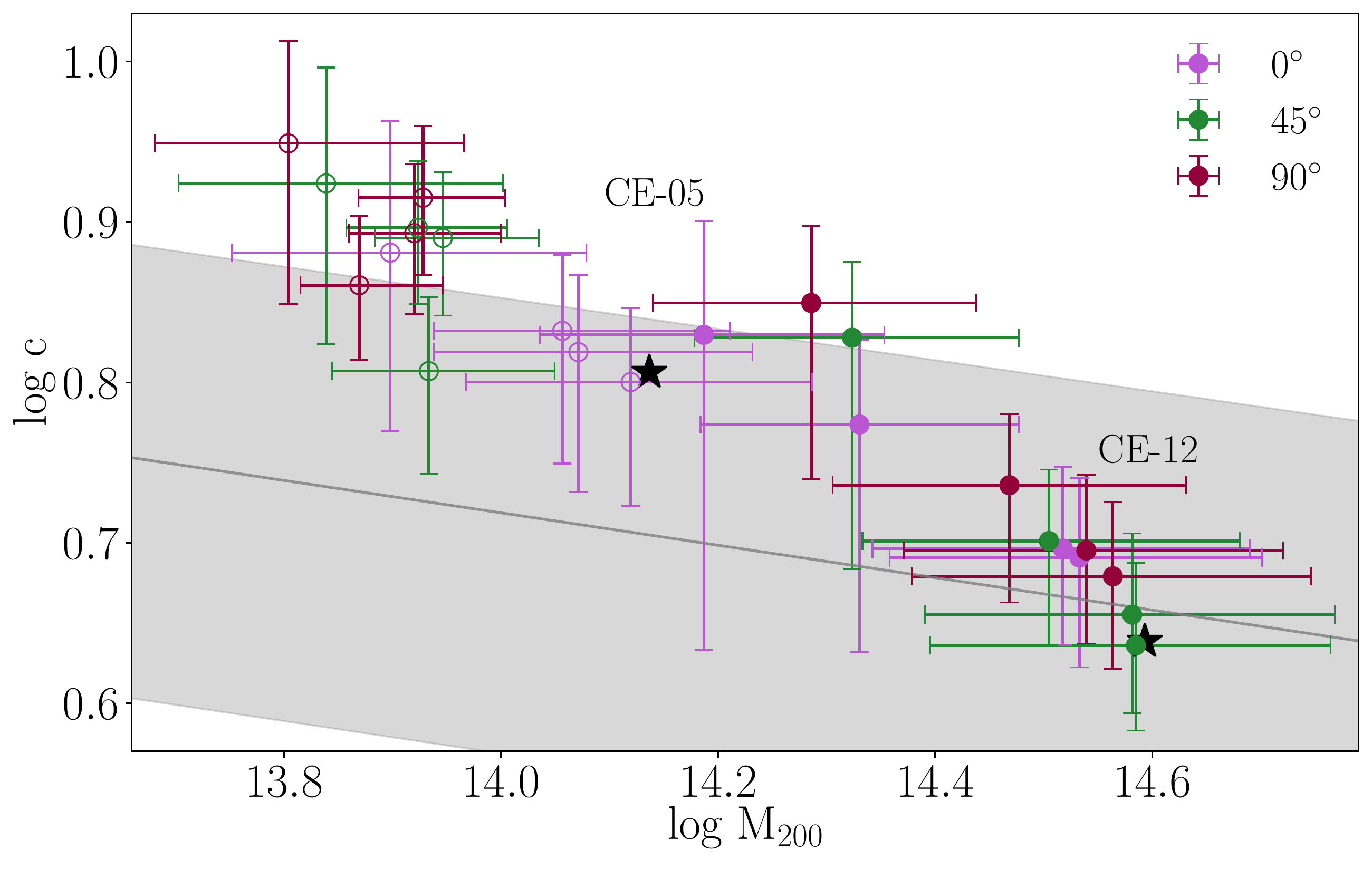}
\caption{\it 
Fitted $M_{200}$ and $c$ values describing the outer collisionless halo in the Jeans model for the simulated clusters.
Stars denote true values of $M_{200}$ and $c$ from CDM-only simulations~\protect\cite{Robertson:2017mgj}.
The left panel corresponds to our primary analysis, with a prior on $M_{200}$ derived from member galaxy kinematics, while the right panel corresponds to a prior centered on the true $M_{200}$ value.
Band shows MCR for $z = 0$, with a width of $\pm{0.15}\,\mathrm{dex}$ in $\conc$. 
}
\label{fig:MCR_plots_sims}
\end{figure}

Finally, in Fig.~\ref{fig:MCR_plots_sims}, we compare the values of $M_{200}$ and $\conc$ from our fits to the true values 
from the corresponding CDM-only simulations for CE-05 and CE-12~\cite{Robertson:2017mgj}. In the Jeans model, these values (denoted by stars) should represent the outer collisionless halo unaffected by self-interactions. 
The left panel shows results from our primary MCMCs, which we recall took priors on $M_{200}$ obtained from galaxy kinematics, as described in Sec.~\ref{sec:MockObsSims}. 
The data points correspond to four lines of sight for each CE-05 (open circles) and CE-12 (closed circles), each for three slit angles. 
We see that our fits systematically pick out somewhat larger values of $c$ and smaller values of $M_{200}$ compared to the CDM-only values, 
as well as show a slight dependence on the slit angle.
The starkest difference is for $M_{200}$ in CE-12, which is due to our prior: the virial masses obtained from galaxy kinematics are biased lower than the true value by as much as a factor of $\sim 2$ (see Table~\ref{tab:stellar_fits}). 

To test the impact of our priors for $M_{200}$ on our results,
we repeated our MCMC analysis using priors centered at the true $M_{200}$ values (with the same width). 
The results are shown in the right panel of Fig.~\ref{fig:MCR_plots_sims}. While the results for CE-05 remain relatively unchanged, those for CE-12 shift closer to the true $M_{200}$, as expected. 
Nevertheless, there remains a systematic trend in our Jeans analyses toward higher concentrations and smaller values of $M_{200}$. We found a related trend for our analysis of the group and cluster observations, obtaining higher $\conc$ with respect to the MCR (Fig.~\ref{fig:MCR_plots_obs}).
However, we again note that the MCR is a CDM-based prediction that may not be representative of SIDM halos (see Sec.~\ref{sec:M200c200}). In any case, we find that our choice of priors for $M_{200}$ has negligible effect on the joint fit for $\sigma/m$ and our results in Eq.~\eqref{eq:sigmam_sim} are virtually unchanged.

\section{Conclusions \label{sec:concl}} 

While self-interactions have long been motivated from dwarf galaxies, it is important to test their effect on systems across all astrophysical scales.
Following the Jeans model for SIDM, we have performed a detailed analysis of the Newman et al.~samples of galaxy groups~\cite{Newman:2015kzv} and massive clusters~\cite{Newman:2012nv,Newman:2012nw}, all of which are strong lensing systems that collectively span $4 \times 10^{13}$--$2 \times 10^{15} \, \Msun$.
From the sample of groups, we obtained a new constraint for $\sigma/m$, the first at an intermediate scale between galaxies and massive clusters. 
We also reassessed the constraint on $\sigma/m$ from massive clusters found in Ref.~\cite{Kaplinghat:2015aga}.
Our analysis has made these constraints more robust by allowing for conservative ranges for various systematic unknowns, which in previous work~\cite{Kaplinghat:2015aga} had either been neglected or subjected to more restrictive assumptions.
These include the unknown stellar mass-to-light ratio $\Upsilon_\star$ and velocity dispersion anisotropy $\beta$, as well as the possible influence of AC on the outer regions of SIDM halos where the collision rate is negligible.

\begin{table}[!b]
\centering
\begin{tabular}{|ll|c|c|c|}
\cline{3-5}
 \multicolumn{2}{c|}{} & $\langle v \rangle \; ({\rm km/s})$ 
 & $\langle \sigma v \rangle /m \; ({\rm cm^2/g \times km/s})$ 
 & $\sigma/m \; ({\rm cm^2/g})$ 
 \\
\hline
Groups & (mean) & $1150$ & $590\pm 370$ & $0.5 \pm 0.2$ \\
& ($95\%$ upper limit) & & $1430$ & $1.1$ \\
\hline
Clusters & (mean) & $1900$  & $300 \pm 150$ & $0.19 \pm 0.09$ \\
& ($95\%$ upper limit) & & $620$ & $0.35$\\
\hline
\end{tabular}
\caption{\it Representative results for $\langle v \rangle$, $\langle \sigma v \rangle / m$ and $\sigma / m$ for the groups and clusters.}
\label{tab:results}
\end{table}

For both groups and clusters, we find a mild preference for a nonzero cross section compared to collisionless CDM.
Table~\ref{tab:results} provides a condensed summary of our results for $\sigma/ m$ and $\langle\sigma v \rangle/m$ for groups and clusters, along with the corresponding values for the velocity $\langle v \rangle$.
In Sec.~\ref{sec:crosssections}, we obtained constraints on $\sigma/m$ separately for cases both without and with AC.
We showed that allowing for AC yields systematically larger values of $\sigma/m$ compared to the case with no AC, although the two cases are consistent at the $1\sigma$ level.
Table~\ref{tab:results} distills the two cases into values that can be used for phenomenological studies without explicit reference to the effect of AC. 
For $\sigma/m$ and $\langle\sigma v\rangle/m$, we quote the mean and standard deviation evaluated from the sum of the joint probability density functions with and without AC.
For $\langle v \rangle$, we take the average of the values from the joint distributions with and without AC.  
Additionally, we quote $95\%$ upper limits on $\sigma/m$ and $\langle\sigma v \rangle/m$, taken from the case including AC since these are more conservative bounds than those without AC.

Our result for galaxy groups represents a new data point on the plane of cross section versus velocity, while for clusters our result is consistent with values around $0.1\,\cmg$ found in previous work~\cite{Kaplinghat:2015aga,Elbert:2016dbb}.
Since typical velocities are larger in more massive systems, our samples of galaxy groups and clusters probe $\sigma/m$ at scales complementary to one another, as well as to rotation curves and other observations on dwarf galaxy scales.
Our results, combined with small scale structure issues for galaxies, are consistent with SIDM with a velocity-dependent cross section that decreases with increasing scattering velocity. 
While the same argument was made previously~\cite{Kaplinghat:2015aga,Elbert:2016dbb}, here we have made this conclusion more robust.
From a particle physics perspective, a velocity-dependent cross section is not ad hoc but rather the general expectation for many well-motivated SIDM models~\cite{Tulin:2017ara}.
For example, a simple dark photon model has the right velocity dependence to explain structure observations across all scales.
Further study of the particle physics implications remains for future work.

In parallel, as a validation of our Jeans-model approach, we have derived and analyzed a sample of mock observations from high-resolution hydrodynamical SIDM simulations of clusters with $\sigma/m = 1\,\cmg$~\cite{Robertson:2017mgj}.
We followed as closely as possible the same methodology as for the groups, constructing the same sets of observables and performing MCMC analyses to determine a joint posterior constraint for $\sigma/m$.
We obtained central values in the range $1.1 $--$1.7\,\cmg$, depending on the assumed position angle for measuring the stellar line-of-sight velocity dispersion profile.
These values are reassuringly similar to the input value $1\; \cmg$, albeit larger by $\sim 0.5$--$1.5$ standard deviations.
We also compared the baryon and dark matter density profiles, as well as other outputs from our fits, to the corresponding quantities obtained from the simulations. 
This comparison yielded generally good agreement but with clear differences between different lines of sight, signaling a breakdown of spherical symmetry.
Moreover, our study is clearly limited in that it is based on only two simulated clusters, one of which is not likely to be selected in strong lensing surveys.
Further analysis with a much larger sample is certainly warranted.

Let us compare our results to other recent limits quoted in the literature for self-interactions on cluster scales.
For relaxed clusters, studies of halo shapes and profiles inferred by strong lensing alone have constrained $\sigma/m < 1\,\cmg$~\cite{Peter:2012jh,Robertson:2018anx} (superseding overestimated limits from two decades ago~\cite{MiraldaEscude:2000qt,Meneghetti:2000gm}).
Here we have obtained stronger limits by using stellar velocity dispersions to constrain the profile at smaller radii, where the cluster is baryon-dominated.
The Jeans model is well-suited to this task since the baryon profile is an input that can be matched directly to observations (up to an overall factor $\Upsilon_\star$), whereas in simulations the stellar profile is an output that cannot be easily tuned to fit a given system.
On the other hand, constraints from galaxy-dark matter offsets in the Bullet Cluster~\cite{Randall:2007ph} and other mergers~\cite{Harvey:2015hha} are often cited as the most stringent limits, but recent work has weakened these limits to $\sigma/m < 2 \, \cmg$~\cite{Wittman:2017gxn}.
A more sensitive probe comes from post-merger offsets of BCGs, which undergo long-lived oscillations (wobbles) in the potentials of cored halos~\cite{Kim:2016ujt,Harvey:2017afv}.
BCG offsets inferred from cluster observations and hydrodynamical SIDM simulations yield a bound $\sigma/m < 0.39\, \cmg~(95\%~{\rm CL})$~\cite{Harvey:2018uwf}, comparable to our results.
It is encouraging that future studies of BCG wobbles with greater statistics may independently corroborate values of $\sigma/m \approx 0.1$--$0.2\,\cmg$ obtained here.

\section*{Acknowledgements}
We thank A.~Muzzin and A.~Newman for helpful discussions. This research was enabled by support provided by Compute Ontario, WestGrid, Calcul Qu\'{e}bec, and Compute Canada, as well as the National Science and Engineering Research Council of Canada. LS is funded by the Deutsche Forschungsgemeinschaft (DFG) through the Emmy Noether Grant No. KA 4662/1-1. 
The work of ST was performed in part at Aspen Center for Physics, which is supported by National Science Foundation grant PHY-1607611. AR is supported by the European Research Council's Horizon2020 project `EWC' (award AMD-776247-6).
The reference list was populated using \textit{filltex}~\cite{Gerosa:2017xrm}.

\appendix

\section{Stellar velocity dispersion\label{sec:losdisp}} 

We summarize the computation of the stellar line-of-sight velocity dispersion $\sigma_{\rm LOS}$, assuming spherical symmetry.
The stellar surface density is
\begin{equation} \label{eq:Sigma}
\Sigma_\star(R) =  \int_R^\infty \diff{r} \frac{2r}{\sqrt{r^2 - R^2}} \, \nu_\star(r) \, ,
\end{equation} 
where $\nu_\star(r)$ is the stellar luminosity density.
The line-of-sight dispersion is computed from the Jeans equation~\cite{1982MNRAS.200..361B}
\begin{equation} \label{eq:Sigmasigma}
\Sigma_\star \sigma_{\rm LOS}^2(R) = 2G \int_R^\infty \diff{r} \frac{\mathcal{F}(r) M_{\rm tot}(r) \nu_\star(r) }{r^{2-2\beta}} \,,
\end{equation}
assuming a constant anisotropy $\beta = 1 - \sigma_t^2/\sigma_r^2$, where $\sigma_{r,t}$ are the radial ($r$) and tangential ($t$) velocity dispersions.
$M_\textrm{tot}$ is the total enclosed mass of stars and dark matter.
We also have
\begin{equation} \label{eq:betaF}
\mathcal{F}(r) = 
\frac{R^{1-2\beta}}{2} \left[ \beta \, B\left(\tfrac{R^2}{r^2} ; \beta + \tfrac{1}{2} , \tfrac{1}{2} \right) - 
B\left(\tfrac{R^2}{r^2} ; \beta - \tfrac{1}{2} , \tfrac{1}{2} \right) + \frac{\Gamma\left(\beta - \tfrac{1}{2} \right) \sqrt{\pi}( 3 - 2\beta)}{2 \Gamma(\beta)} \right], 
\end{equation}
where $B$ is the incomplete beta function~\cite{Cappellari:2008kd}. 
Eq.~\eqref{eq:betaF} reduces to $\mathcal{F}(r) = \sqrt{r^2 - R^2}$ for $\beta=0$.
The limits $\beta = 0, 1, -\infty$ correspond to isotropic, radial, and circular orbits, respectively.

Before taking the ratio of Eqs.~\eqref{eq:Sigma} and \eqref{eq:Sigmasigma}, we must account for seeing and the finite slit geometry in order to connect with observations.
Including both effects~\cite{Sand:2003bp}, the quantity to be compared to observations is
\be
\sigma_{\rm LOS}^2 = \frac{  \int \diff{A}  \widetilde{ \Sigma_* \sigma^2_{\rm LOS}} (R)  }{ \int \diff{A}  \widetilde{ \Sigma_*}(R) } \, .
\ee
We correct Eqs.~\eqref{eq:Sigma} and \eqref{eq:Sigmasigma} for seeing using a Gaussian PSF with width $\sigma_{\rm PSF}$ according to the formula
\begin{equation} 
\widetilde{f}(R) = \int_0^\infty \diff{R^\prime} R^\prime f\left(R^\prime\right) I_0\left( \frac{R R^\prime}{\sigma_{\rm PSF}^2 }\right)
\exp\left( \frac{ R^2 + R^{\prime 2}}{2 \sigma_{\rm PSF}^2 }\right) \, ,
\label{seeing}
\end{equation}
where $I_0$ is a modified Bessel function~\cite{1982MNRAS.200..361B}. 
Lastly, the integral $\int \diff{A}  = \int_{R_{\rm min}}^{R_{\rm max}} \diff{x} \int_{-w/2}^{w/2} \diff{y}$ provides a spatial average over the slit width $w$ and the range $R_{\rm min} < R < R_{\rm max}$ for a given projected radius bin.

Lastly, the slit geometry is circularized to correct for ellipticity on the plane of the sky. This procedure matches an ellipse with semi-major and -minor axes lengths $a,b$ to a circle of radius $\sqrt{ab}$.
A point at physical projected radius $R$ and relative angle $\phi$ between the slit and major axis is circularized to radius $R_{\rm circ} = k R$. 
The correction factor $k = \sqrt{ q \cos^2\phi + q^{-1} \sin^2\phi}$ is used to circularize $R_{\rm min},$ $R_{\rm max}$, defining the extent of the bin along the slit, where $q=b/a$ is the axis ratio. 
The slit width $w$ is circularized in the orthogonal direction and one must replace $\phi \to \phi + \pi$ for $k$. 
For a slit aligned (perpendicular) with the major axis, we have $k=\sqrt{b/a}$ ($k = \sqrt{a/b}$).

\section{Enclosed mass profiles for clusters  \label{app:M2D_profile}} 

\begin{figure}[t!]
	\centering
	\includegraphics[width=0.49\textwidth]{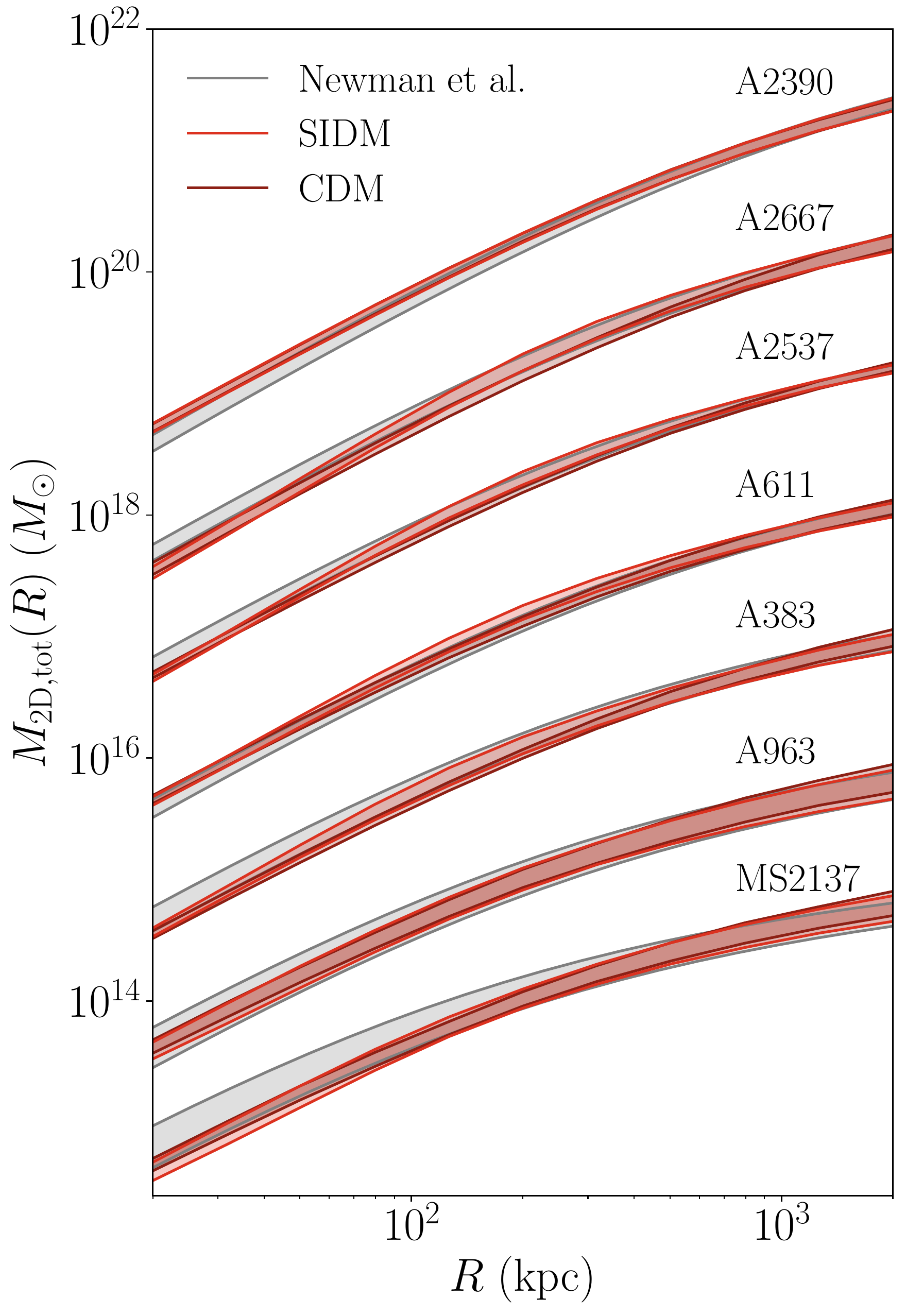}
	\caption{\it Our SIDM (red) and CDM (purple) results for $M_{\rm 2D, tot}(R)$, the total 2D mass enclosed within a projected radius $R$, for the sample of clusters. These results are consistent with profiles obtained by Newman et al.~\protect\cite{Newman:2012nv} by lens reconstruction analysis (gray). Note profiles are offset from one another by $+1\,\mathrm{dex}$.}
	\label{fig:M2Dtotal}
\end{figure}

Since our analysis takes a simplified approach for fitting strong and weak lensing observations for clusters, we compare the projected mass profiles from our fits to those obtained by full lens reconstruction by Newman et al.~\cite{Newman:2012nv}.
Fig.~\ref{fig:M2Dtotal} shows $M_{\rm 2D,tot}(R)$, the 2D projected total mass enclosed within radius $R$ for each cluster, each subsequently shifted upward by $+1$\,dex from MS2137.
Results from our SIDM (CDM) fits, without AC, are shown by the red (purple) bands, whose width denotes the $1\sigma$ range.
For comparison, the gray band corresponds to the Newman et al.~\cite{Newman:2012nv} results for these systems from strong and weak lensing only (i.e., without stellar kinematics).
The width of the band corresponds to the $1\sigma$ allowed range for ($M_{200},c$) for a spherically-averaged NFW profile for the total mass profile of the cluster halo and BCG.
It is reassuring that our SIDM and CDM fits are consistent with the results of Ref.~\cite{Newman:2012nv}.

\vspace{2.5cm}

\bibliography{mega_dm_bib}
\bibliographystyle{apsrev4-1}
\end{document}